\documentclass[a4paper,12pt,onehalfspacing,headrules,twoside]{report}
\usepackage[latin1]{inputenc}
\usepackage{times}
\usepackage{latex8}
\usepackage{longtable}
\usepackage{breakcites}
\usepackage{slashbox}
\usepackage{multirow}
\usepackage[T1]{fontenc}
\usepackage[french]{babel}
\usepackage{pst-all}
\usepackage{pst-node}
\usepackage{pst-tree}
\usepackage{graphicx}
\usepackage{amsfonts}
\usepackage{amsmath}
\usepackage{amssymb}
\usepackage{fancyhdr}
\usepackage[ruled,vlined,english,titlenumbered]{algorithm2e}

\newcommand{\twlrm}{\fontsize{9}{8.3pt}\normalfont\rmfamily}
. scaled \magstephalf

\fancypagestyle{plain}{ 
\fancyhead{} 
}
\renewcommand{\chaptermark}[1]{\markboth{#1}{}}

\begin{document}
\vspace*{-1.5cm}
\thispagestyle{empty}
\begin{center}
\textsc{\textbf{Universit\'e De Tunis El Manar}}\\
\textsc{\textbf{Facult\'e Des Sciences De Tunis}}
\end{center}
\vspace{-0.3cm}
\begin{center}
\begin{figure}[htbp]
    \centering
\includegraphics[scale=0.4]{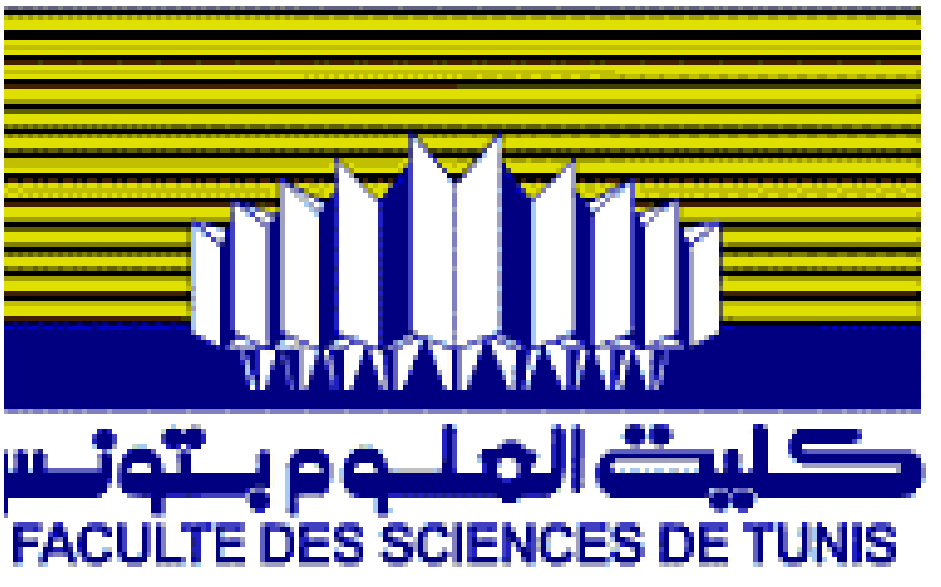}
\end{figure}
\end{center}
\begin{center}
\vspace{-1.cm}
\textsc{\textbf{\'Ecole Doctorale En Mathématiques, Informatique, Sciences Et
Technologies De La Matière}}

\bigskip

\mbox{}

\bigskip

\mbox{}

\bigskip

{\LARGE \textbf{Replication in Data Grids: Metrics and Strategies}}

\bigskip

\mbox{}

\bigskip

{\Large \textbf{Mémoire de Synthèse}} \\

\mbox{}

\bigskip

présenté et soutenu le 15 juillet 2017\\\mbox{}\\ en vue de l'obtention de l'

\bigskip

\textbf{\LARGE{Habilitation Universitaire en Informatique}} \\

\mbox{}

\bigskip

par

 \bigskip

 {\Large\textbf{Tarek \textsc{Hamrouni}}}\\
  {\large Docteur en Informatique, Maître-Assistant à\\l'Institut Supérieur des Arts Multimédias de La Manouba}

 \bigskip

  \mbox{}

 \bigskip

 \bigskip

  \mbox{}

 \bigskip

\mbox{}

 \bigskip

devant le jury composé de :

 \bigskip

\begin{tabular} {llll}
\textbf{M.} &\textbf{Ezzeddine \textsc{Zagrouba}} & \textbf{Université de Tunis El Manar} & \textbf{Président}\\
\textbf{M.} &\textbf{Sami \textsc{Bhiri}} & \textbf{Université de Monastir} &\textbf{Rapporteur}\\
\textbf{M.} &\textbf{Khalil \textsc{Drira}} & \textbf{Université de Toulouse} &\textbf{Rapporteur}\\
\textbf{M.} &\textbf{Faouzi \textsc{Ben Charrada}}  & \textbf{Université de Tunis El Manar} &\textbf{Examinateur}\\
\textbf{Mme.} &\textbf{Amel \textsc{Grissa Touzi}}  & \textbf{Université de Kairouan} &\textbf{Examinatrice}\\
\end{tabular}

\end{center}

\newpage

\thispagestyle{empty}

\mbox{}

\newpage

\thispagestyle{empty}

\mbox{}

\vspace{5.cm}

\parbox{16cm}{\hspace{6.cm}
\textit{To the memory of my exemplary father Abdelwaheb,}}
\parbox{16cm}{\hspace{6.4cm}
\textit{for all he did for me...}}

\vspace{1cm}

\parbox{16cm}{\hspace{6.cm}
\textit{To my dear mother Oum-Aziz,}}
\parbox{16cm}{\hspace{6.4cm}
\textit{To my beloved wife Hanène,}}
\parbox{16cm}{\hspace{6.4cm}
\textit{for their support...}}

\vspace{1cm}

\parbox{16cm}{\hspace{6.cm}
\textit{To my lovely daughter Mayar \textsc{(}Mayoura\textsc{)},}}
\parbox{16cm}{\hspace{6.4cm}
\textit{for enjoying my life...}}

\newpage

\thispagestyle{empty}

\newpage
\thispagestyle{empty}

\mbox{}

\newpage

\thispagestyle{empty}
\begin{center} {\Large \textbf{Acknowledgements}}
\setcounter{footnote}{0}
\markboth{Acknowledgements}{Acknowledgements}
\end{center}

\bigskip

\bigskip

I would like to thank the many people that contributed to the realization of
this habilitation thesis.

\bigskip

I would like to thank my habilitation thesis committee. Foremost, I express my gratitude to Professor Ezzeddine \textsc{Zagrouba} to have agreed to chair the evaluating committee of my habilitation thesis.

\bigskip

I would like to express my sincere thanks to Professor Sami \textsc{Bhiri} and Professor Khalil \textsc{Drira} who have kindly volunteered their time and accepted to review the manuscript. I am also thankful to Professor Faouzi \textsc{Ben Charrada} and Professor Amel \textsc{Grissa Touzi} for agreeing to participate to the thesis committee.

\bigskip

I would sincerely like to thank the researchers without whom none of this would have been possible:
\begin{itemize}
  \item Professor Faouzi \textsc{Ben Charrada}, for valuable advices and all the useful discussions and clever ideas during key steps of the habilitation thesis.
  \item Professor Sadok \textsc{Ben Yahia}, for his guidance and assistance during my 
      first research supervisions.
  \item My master and Ph.D. students for their devoted efforts.
  \item People who kindly provided the source codes of their strategies which helped in the evaluation of our contributions.
\end{itemize}

\bigskip

During my habilitation thesis, I have been an Assistant Professor at the Higher Institute of Multimedia Arts of La Manouba \textsc{(}ISAMM\textsc{)}. I have carried out my research work at the Department of Computer Sciences of the Faculty of Sciences of Tunis \textsc{(}FST\textsc{)}. In this respect, I would like to thank all the ISAMM and FST employees for their kindness and support during these years.

\bigskip

Last, but not least, I would like to express my gratitude to my family. Special thanks go to my mother and my wife for consistently and faithfully supporting me through all this period and encouraging me to go further in my research career.

\bigskip

This habilitation thesis is dedicated to the memory of my father, for the researcher that he was and for the love of research that he inculcated to me.

\newpage
\mbox{}
\thispagestyle{empty}
\newpage

 \pagenumbering{roman}
\thispagestyle{empty} \tableofcontents
\newpage
\thispagestyle{empty} \listoffigures
\addcontentsline{toc}{chapter}{List of Figures}
\newpage
\mbox{}
\thispagestyle{empty}
\newpage
\thispagestyle{empty} \listoftables
\addcontentsline{toc}{chapter}{List of Tables}
\newpage
\mbox{}
\thispagestyle{empty}
\newpage
\thispagestyle{empty} \listofalgorithms
\addcontentsline{toc}{chapter}{List of Algorithms}
\newpage

\thispagestyle{empty}

\mbox{}

\newpage
 \linespread{1.3}
 \normalfont

\pagenumbering{arabic}
\chapter*{Introduction}
\addcontentsline{toc}{chapter}{Introduction}
\markboth{Introduction}{Introduction}
\setcounter{footnote}{0}

Data-intensive applications in the era of Big Data \cite{BigData2014Ref2,BigData2014Ref1} are becoming increasingly prevalent in domains of scientific and engineering research such as high energy physics, earth science, bioinformatics, data mining, astronomy, to quote but a few. 
In this context, data grids emerge as a suitable solution for these applications. Data grid is an integrating architecture that allows connecting a collection of hundreds of geographically distributed computers and storage resources located in different parts of the world to facilitate the sharing of data and resources \cite{fosterBook2003,datagrid2000}. Data grids primarily deal with providing services and infrastructure for distributed data-intensive applications that need to access, transfer and manage massive data sets stored in distributed storage resources. In this kind of dynamic and large scale environment, a lot of challenges revolve around data management and transfer \cite{Allcock2002,Vazhkudai2001}.\\

The grid has then for objective to allow users exploiting remote resources no matter where they are located to execute jobs requiring locally unavailable resources. Each grid site then supplies computing and storage resources to the other sites. A job executed in a given site may require data which can be found in the site where it is executed as it may be found in other sites. In this regard, the effective data management is one critical issue in data grid systems and involves many challenges. One way to effectively cope with these challenges is to rely on the replication technique. Replication consists to create multiple copies of the same data \textsc{(}\textit{aka} replicas\textsc{)} in several storage resources. Hence, data availability is increased since data are stored at more than one site. In contrast, if data are not replicated in several grid sites, all data requests issued from grid users or grid jobs must wait at a single site. The grid system is then challenged by many problems such as the increased risk of failures, the overloading of popular sites, and access latency.\\

In this context, 
data replication clearly scales up the performance by reducing remote access delay and mitigating single point of failure \cite{Boru2015}. In addition, it helps overcoming long wide-area data transfer latencies by keeping data close to locations where queries are originated \cite{Allcock2002}. Indeed, through replication, data grid can achieve high data availability, improved bandwidth consumption, and better fault tolerance, while ensuring a certain distribution transparency \cite{DistributedChallenges}. In this regard, replication is also a popular technique in many distributed systems \cite{Goel06datareplication,ReplicationTaxonomy2016} such as distributed databases systems \cite{DBreplicationsurvey2000}, mobile systems \cite{surveymobile2008}, cloud systems \cite{pCloudSurvey,surveyCloudJNCA2016}, Peer-to-Peer \textsc{(}P2P\textsc{)} systems \cite{pReplicationP2P,pReplicationStrategiesP2P} as well as Content Delivery Network \textsc{(}CDN\textsc{)} systems \cite{pCDN1,pCDN2}, to quote but a few.\\

Most of the works in the literature dedicated to the proposal of replication strategies in data grids assume that the data is read-only \cite{surveyFGCS2012,surveyFGCS2013,surveyCloudJNCA2016,MoHa2015.1}. Replication can indeed greatly improve the data grid performance in case of read-only data \cite{Goel06datareplication}. However, when data updates are allowed, the benefits of replication can be neutralized by the overhead of maintaining consistency among multiple replicas \cite{4459673,consistency2017CCPE}. Indeed, data replication implies that consistency and availability cannot be simply combined in the presence of network partitions \textsc{(}the CAP theorem\textsc{)} \cite{Gilbert:2002:BCF:564585.564601}.
In all our contributions, we mainly focus on the scenario dealing with the read-only data, \textit{i.e.}, without consistency management.\\

A data replication strategy involves several issues. We summarize the most critical issues that the researchers have agreed about in the following:
\begin{itemize}
 \item Which data must be replicated?
 \item Where to place the replicas \textsc{(}replica placement\textsc{)}?
 \item From which site a requested file must be transferred in the presence of multiple replicas \textsc{(}replica selection\textsc{)}?
  \end{itemize}
Depending on the considered issues, intensive research has been conducted on developing data replication strategies \cite{surveyFGCS2012,Grace,surveyFGCS2013,TosCC2015}.\\

On the other side, there has been recently significant interest on using data mining in grids. Data mining is defined as the process of uncovering new meaningful, previously unknown knowledge and information from large amounts of data. 
The emergence of grid technology and the increasingly complex nature of data mining applications have indeed led to a new synergy of data mining and grid \cite{mining_grid_Foster07}. In this respect, the concept of data mining grid allows data mining process to be deployed in a grid environment. The grid provides indeed an effective computational support for distributed data mining applications \cite{distributedDM2007}. On the other side, mining grid data is an interesting research field which aims at analyzing historical data that could be obtained from grid system, with data mining techniques in order to efficiently discover new meaningful knowledge to enhance grid systems in many areas \cite{SMDVPD08} such as resource management \cite{Khanli2011703}, scheduling \cite{Asgarali2009}, failure detection \cite{1652162}, data transfer \cite{dataaccessprediction2014}, etc. Data replication strategies can then benefit from data mining techniques for performing several key tasks like discovering data file correlations, predicting the future file usage based on past history as well as the future value of a parameter based on related ones \cite{EAAI2016}.\\

\bigskip


In this report, we present a synthesis of our main contributions to the field of replication strategies in data grids. Our contributions can be divided into two main axes which are described as follows.


In the first one, we mainly focus on the placement of replicas over grid sites since a given placement may have not only an impact on the current execution of the grid, but also on the next executions. Indeed, the strategy invoked to place replicas through the sites of the grid constitutes one of the main factors that can have a great influence on the grid performance for a long time \cite{MAGS2015}. In this respect, it is important to stress on the fact that the replicas of a given file can be distributed in several different ways through the sites of the grid, which allows generating several possible distributions. The quality of the generated distribution has then a great impact on the obtained performances. 

However, almost all placement strategies mainly focus on the current execution of the grid and do not take into account future ones \cite{MAGS2015}. This important point motivated us to study in-depth the impact of a replication strategy on the replicas distribution as well as the opposite view, \textit{i.e.}, the impact of a distribution on the results of a replication strategy. Indeed, a strategy may be efficient only for a short period while its negative impact through placing replicas in a given set of sites will be observed in the future executions.

In the literature, researchers have mainly focused on what the replication strategy can give as services in order to accelerate the current execution or minimize its cost \cite{surveyreplication2013}. In our proposition, we are going to highlight the importance of the fact that the strategy may offer services not only for the current jobs executions but also for the forthcoming grid users. Indeed, a strategy can leave badly placed replicas than that it found before its invocation, as it can lead to a better replicas distribution. This will either negatively or positively affect the performance of the grid. Hence, the obtained results using classical metrics \textsc{(}execution time, consumed bandwidth, etc.\textsc{)} will not allow an ``objective'' evaluation of a strategy. Indeed, these results are influenced by the distribution on which the strategy is invoked. A strategy then ensures a high quality of replicas distribution when:\\\hspace*{0.5cm}\textsc{(}$i$\textsc{)} Many requests are satisfied locally, \textit{i.e.}, they are satisfied by the same site where \hspace*{1.cm}the request is generated.\\\hspace*{0.5cm}\textsc{(}$ii$\textsc{)} Links with high bandwidth are preferentially used when accessing data remotely.\\\hspace*{0.5cm}\textsc{(}$iii$\textsc{)} The most available sites are the most used.\\
Satisfying these complementary three points by a replication strategy helps to reduce execution time and optimize resource consumption for the current but also for the forthcoming uses of the grid. Moreover, as also highlighted in \cite{Vrbsky2013}, replication strategies that decrease the file access times by accessing local files, while not requiring an increase in data storage, are needed for energy efficiency.\\

We focus in this report on the following main contributions w.r.t. this axis:
\begin{itemize}
  \item An overview of replication strategies mainly from the viewpoints of the considered parameters in their associated steps as well as the used metrics in the literature for their evaluation.
  \item A study of the impact of placement strategies on data grid performance which motivated the analysis of the effect of the replicas distribution quality on the performance results of replication strategies.
  \item The proposal of new evaluation metrics dedicated to the evaluation of the distribution quality. Having each its own properties, they also have for common purpose to highlight the ability of a replication strategy to place replicas in strategic locations and, hence, its influence on the future uses of the grid.
  \item The setting of an ``objective'' evaluation of replication strategies which is based on a beforehand assessment of the distribution quality. Noteworthily, a large part of the evaluation metrics of replication strategies is strongly influenced by the quality of the replicas distribution whose correct assessment will have a key role in the contextualization of the results obtained through these metrics.
\end{itemize}

The contributions associated to this axis are mainly published in five journal articles \cite{CC2016,JNCA2016,JNCA2015,MAGS2015,IJWET2016} and two conference papers \cite{3PGCIC2014,CoopIS2016}.\\

\bigskip


The second axis is mainly dedicated to exploiting results of data mining techniques to enhance performances of replication strategies. Indeed, such techniques allow for example strategies to rely on sets of files as granularity level for replication instead of a single file. Moreover, they can also help predicting the value of some parameters based on those of others. Even though data mining has been applied in numerous areas and sectors, the application of data mining to replication in the context of data grids is still limited. Only few works, when comparing their number with that of all replication strategies, have indeed used data mining techniques to explore file correlations although the strength of data mining approaches in real-life applications.

Among data mining-based replication strategies, a large part relies on frequent patterns extraction and association rule generation. Nevertheless, two main problems hamper the adequate exploitation of the extracted knowledge from these sets of patterns. The first is related to the quality of the extracted patterns which do not necessarily reflect true correlations between mined attributes \textsc{(}files in the case of replication strategies\textsc{)}. The second problem is often related to the huge quantity of the extracted patterns. This motivated researchers to only mine concise representations of pattern sets \cite{IDA2012} using for example the Formal Concept Analysis \textsc{(}\textsc{FCA}\textsc{)} mathematical foundations \cite{ganter99}.\\

To go further towards reducing the number of mined patterns while retaining the more informative ones, many works in the literature propose to integrate the correlation measures within the mining process \cite{Brin97,ccmine_Kim,comine_Lee,Omie03,Xiong06hypercliquepattern}.
Correlated pattern mining is then shown to be more complex but more informative than the traditional pattern mining. In fact, correlated patterns offer a precise information about the degree of apparition of the attributes composing a given patterns \cite{SegondB12a}. In addition, the use of condensed, \textit{aka} concise, representations of the whole correlated patterns allows reducing the number of the extracted patterns while preserving pertinent knowledge \cite{IDA2012}.

From the view point of correlated pattern mining, some of our work was dedicated to the proposal of concise representations of frequent \cite{DS2010} as well as rare \cite{IDA2015} correlated patterns using a correlation measure such as \textit{all-confidence} or \textit{bond} \cite{Omie03}. This motivated us to exploit the obtained theoretical results in order to apply them in the context of replication strategies in data grids. We then propose a new dynamic replication strategy which takes into account file correlations. Maximal frequent correlated pattern mining is used to find groups of correlated files. Replication is then based on the granularity of a set of correlated files instead of a single file. The proposed strategy consists of three main steps: storing file access history and converting it into an extraction context, applying maximal frequent correlated pattern mining algorithm and, finally, performing replication and replacement using the extracted knowledge.\\

With respect to this axis, we mainly concentrate on the following contributions listed below:
\begin{itemize}
  \item The study of the strengths and the drawbacks of the main replication strategies based on data mining techniques and how these latter are applied in this context.
  \item The proposal of a new guideline to data mining application in the context of data grid replication strategies. This guideline describes finely directive lines to be followed when designing a new replication strategy based on the results of data mining techniques. Also, it describes in detail the main steps, the decisions to be made, the specificities and the constraints to be taken into account stemming from both data grid and data mining contexts.
  \item The proposal of a new algorithm for mining maximal frequent correlated patterns. The input of this algorithm is obtained through a preliminary step focusing on how to adapt the required grid concepts to the data mining algorithm.
  \item The design and the implementation of a new replication strategy based on a data mining technique, and more precisely correlated patterns.
\end{itemize}

Our contributions in this axis are mainly published in four journal articles \cite{TSI2012,IDA2015,JSS2015,EAAI2016} and four conference papers \cite{DS2010,PAKDD2012,ICCS2015,PDCAT2014}.

Please note that for the sake of brevity, we omit in the dedicated chapter several details w.r.t. our study of the state of the art strategies based on data mining techniques as well as the set up guideline. We then refer interested readers to our work published in \cite{ICCS2015,EAAI2016}. In addition, the description of our contributions mainly focusing on the mining of \textsc{(}reduced sets\textsc{)} of \textsc{(}correlated\textsc{)} patterns is omitted and can be found mainly in \cite{DS2010,TSI2012,PAKDD2012,IDA2015,IDA2012,IJAIT2014}.\\

\bigskip

The last part of the report is dedicated to a summary of our contributions and a discussion on future work.

\newpage

\thispagestyle{empty}

\mbox{}

\newpage

\chapter{Distribution Quality-based Evaluation of Replication Strategies}\label{chapter_quality_of_distribution}
\setcounter{footnote}{0}

\section{Introduction and motivations}

In data grids, several replication strategies \cite{surveyFGCS2012,surveyreplication2013,Grace,EAAI2016,surveyFGCS2013,TosCC2015} have been proposed in the literature in order to overcome the encountered difficulties, and to ensure that the decision of replication was made to the appropriate file, at the right time and in the best location. 

Replication strategies can then be categorized into two main classes: static and dynamic. In static strategies, any created replica is kept in the same place until its lifetime expires or the user deletes it manually after a long period of services. In contrast, dynamic strategies create and delete replicas according to changes in the environment of the data grid. As the environment is dynamic, dynamic strategies are more appropriate for these systems. For example, any site can join or leave the network at any time. As a consequence, it is in general crucial to take into account this dynamicity aspect.\\

Dynamic strategies can be classified according to the following complementary criteria: \label{criteriaForReplicationStrategiesClassification}
\begin{itemize}
\item \textbf{Periodicity:} this criterion specifies when the replication strategy is triggered, at each file request or after a given period of time. We then distinguish non periodic strategies \cite{Chettaoui2014} \textsc{(}\textit{aka} unconditional replication strategies \cite{MisYong2008}\textsc{)} which perform replication for every file request from periodic replication strategies \textsc{(}\textit{aka} conditional replication strategies\textsc{)} which are triggered at each given period. Note that the period can be either static or dynamic. In the static way, the period is constant and can be defined through a time period or by a fixed number of jobs. In the dynamic way, several criteria can be taken into consideration in order to adapt the  duration of the period to the behavior of the grid.

\item \textbf{Nature of decision:} this criterion distinguishes between centralized and decentralized replication strategies. In the first case, there is a single entity which is responsible of managing the replication process \textit{e.g.} replica creation, replica placement, replica replacement, etc. This can lead to the overload of the central site which represents a single point of failure. In the case of decentralized replication, multiple entities such as grid sites and grid users can make decision of replication. Hence some synchronization is involved in order to provide better results. The most important advantage of a decentralized decision is that there is no single point of failure. However, unnecessary replications can be carried out since replication decisions are based on partial information.

    \item \textbf{Grid topology:} there are several grid topologies on which the design of a strategy can be performed. We then distinguish for example strategies dedicated for hierarchical grids and other dedicated for P2P or hybrid grids.

     \item \textbf{Storage space capacity:} the assumption about the size of the available storage space has also a strong impact on the applicability of the replication strategies in the real case. In this regard, some replication strategies assume that each storage space contained in a given site has a limited storage capacity. In this case a replacement strategy is invoked when there is no available space for storing a new replica. In the other case, strategies consider this capacity as unlimited.



     \item \textbf{Evaluation metrics:} the strategies can also be classified according to the metric \textsc{(}or set of metrics\textsc{)} they attempt to optimize. Indeed, we can find a strategy dedicated to the reduction of bandwidth consumption, while another tries to increase the data availability, etc.
\end{itemize}

\begin{figure}[h]
\centering
\includegraphics[scale=0.4]{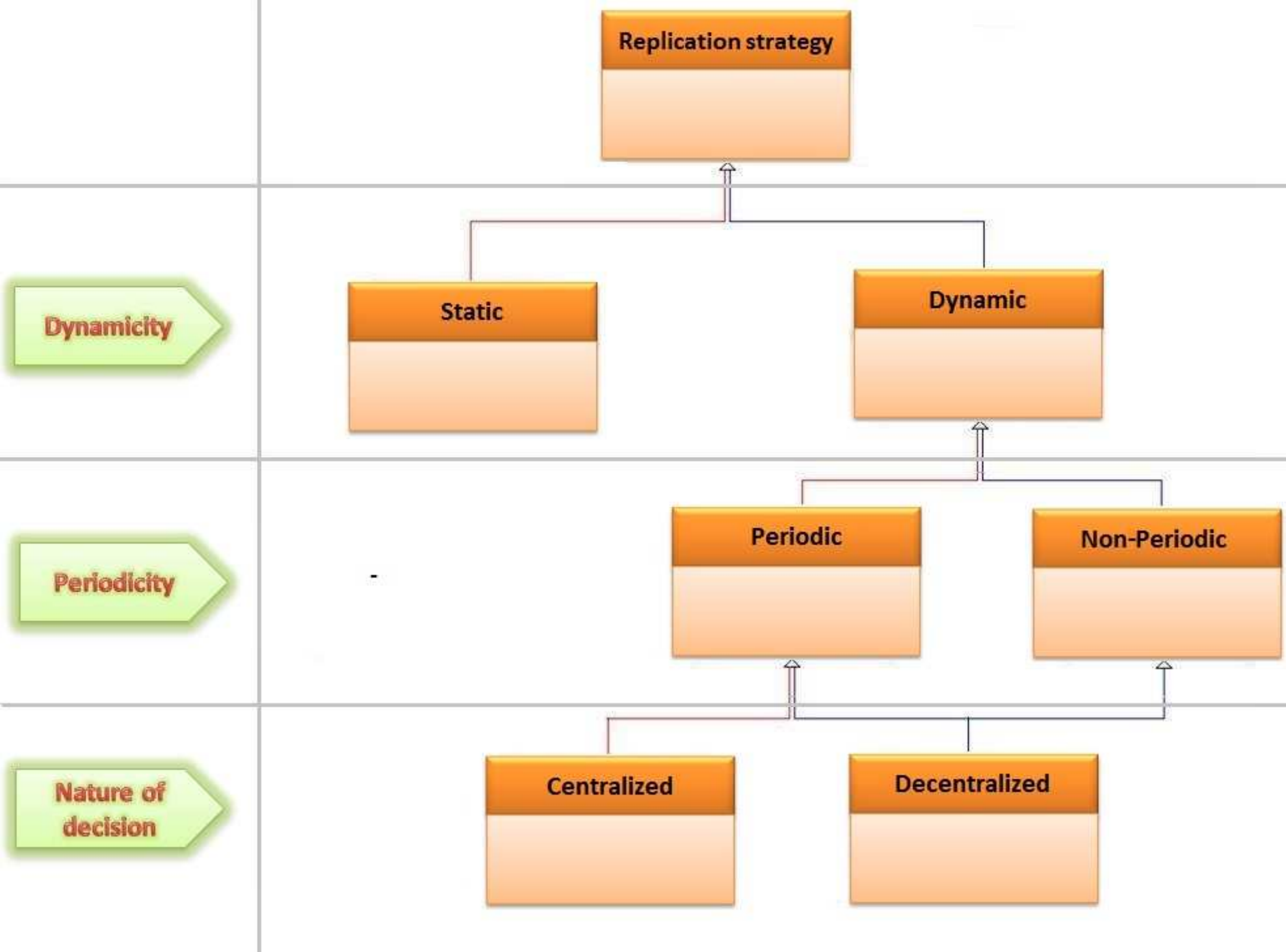}
\caption{\label{Classification des strategies de replication} Classification of replication strategies}
\end{figure}

Figure \ref{Classification des strategies de replication} \cite{JNCA2015} proposes a rough classification of replication strategies with respect to the three main criteria which are the dynamicity, the periodicity and the nature of decision. For a description of the main replication strategies, interested readers are referred to several survey papers in the literature like \cite{surveyFGCS2012,surveyreplication2013,Grace,EAAI2016,surveyFGCS2013,TosCC2015}. In particular, interested readers are referred to \cite{JNCA2015,EAAI2016} for an overview of the replication strategies proposed in the literature and the parameters they use as well as a description of the evaluation metrics of replication strategies.

In this respect, one can evaluate the effectiveness of these strategies through several evaluation metrics, for example, total job time, the consumption of bandwidth, the number of replications performed, the percentage of the processors use, the storage space used, etc. \cite{JNCA2015}. It is, however, important to note that there is not a possible unique metric covering all the aspects of the evaluation of replication strategies.\\

Once a strategy achieved its execution, replicas can have either an interesting distribution over grid sites or not. This will positively or negatively affect the results of next jobs executions. We then focus on the quality of the replicas distribution obtained by the replication strategy. This allows to assess the power of the strategy towards the improvement of data availability and the reduction of the total job time as well as the resources consumption. 
Moreover, the obtained performances of replication strategies are influenced by the distribution quality of replicas. Hence, if we can correctly evaluate it, we can get a priori idea about the performance results that will be obtained using evaluation metrics.\\ 

To the best of our knowledge, there is no previous research that has been made dealing with the distribution quality of replicas in grid sites. Worth noting, there are some similar researches in other domains. For example, in Marketing, researchers have focused on this aspect \cite{Maranzana,url}. This allowed them to obtain higher rates of product availability, and consequently an increase in customer satisfaction. In the electricity domain, several works have focused on the decrease of the total power loss and the improvement of the power quality of distribution systems. This is carried out through the use of distributed generators while identifying their optimal number and their suitable locations in the system \cite{electricity}. On another side, several works in data mining were focused on assessing the sparsity/density of the data from which patterns will be mined \cite{IS2013}. This indeed helps in adopting the more adequate algorithm for the mining task according to this key information.\\
%
%

An in-depth analysis of the replica placement strategies \cite{MAGS2015} allows to note that the choices made by placement strategies not only have an impact on the current execution of the grid but also on the next executions since their impact will last for a while. Unfortunately, the impact of placement strategies on the future executions of the grid has not yet been studied in great depth. Indeed, as discussed in \cite{pTemporalLocalityDistrSyst} and \cite{pTemporalLocalityGrid}, changes in the behavior of the distributed systems do not come all at once but rather gradually. This is also supported by the temporal locality notion \cite{pTemporalLocality,Ranganathan01identifyingdynamic}. In this chapter, we show that different distributions of file replicas have distinct impacts on grid performance. This proves the importance of assessing distribution quality towards a more precise evaluation of replication strategies performances. 

As a matter of fact, the replicas distribution quality can be seen from several angles of view and according to several criteria like the data access time, the data storage cost, the load balancing between grid sites, etc. However, our vision on the distribution quality is interested with the ability of the replicas distribution to maximize the number of local accesses and minimize the cost of remote accesses. An efficient strategy w.r.t. the distribution quality point of view is then the one that offers many local accesses, and only few remote ones. In addition, these remote accesses should not be costly. In this way, increasing the distribution quality will constitute a shortcut for the improvement of the system performance in terms of data availability, execution time and network traffic in the short-term as well as in the long-term future \cite{CoopIS2016}.

In this respect, we propose several metrics for assessing the quality of the replicas distribution. Each one of these metrics has its proper characteristics compared to the other proposed ones. Indeed, a first one is mainly dedicated to the cost of remote accesses to replicas, while a second one is proposed in order to analyze the effect of replications on the distribution of replicas. A third metric is also designed in order to take into consideration both local and remote accesses.
In the general case, our evaluation of the distribution quality of replicas on grid sites takes into account file access history, bandwidth between sites, number of local and remote reads of each replica, and probability of site failure.


A fundamental idea of this work is the correction of the results of evaluation metrics by taking into consideration the initial distribution quality. The third metric we propose contributes to the evaluation of how much the replicas distribution was at the advantage/disadvantage of the strategy performances. This offers a reliable evaluation process of the results that will be independent of the effect of the distribution pre-existing before a strategy is executed.



\section{Influence of replicas distribution on the evaluation results}\label{Sectioninfluence}

The main objective of this section is to highlight the importance of assessing the quality of replicas distribution as a main parameter for evaluating the performances of replication strategies \cite{3PGCIC2014,JNCA2015}.

\subsection{Proof of the influence of replicas distribution on the evaluation results}\label{influence}



In order to highlight the importance of taking into consideration the distribution quality factor, we perform a series of experiments on the DR2 \cite{DR2} and the Periodic Optimiser \cite{PeriodicOptimiser} strategies. The choice of these strategies is argued by the parameters taken into consideration by each strategy, which have a great ability to influence the placement of replicas in grid sites like the availability of sites, the number of requests and the bandwidth.

In this regard, the OptorSim simulator \cite{CameronEtAlOptorSim} is used under the grid configuration given by the CMS testbed, created for the challenge of managing a huge volume of data. This simulator is adopted by most of the works dedicated to replication strategies \cite{surveyFGCS2012,Grace}. OptorSim is a simulation package written in Java language, used to simulate the data grid structure and test job scheduling and replication strategies.\\

For each strategy, we start from two different distributions $D1$ and $D2$ and under the same conditions that is to say, for the same set of jobs, the same files and the same number of replicas for each file. However, when comparing both distributions, the replicas are distributed in different ways, \textit{i.e.}, on different sites. We evaluate both replication strategies using two main metrics frequently relied in the literature and which are:
\begin{itemize}
 \item \textbf{The response time \textsc{(}RT\textsc{)}:} it is the
time elapsed since a job sends a request for files until it
receives all the requested files. If the requested file is
stored locally, the response time is very small and can
be omitted.

\item \textbf{Effective Network Usage \textsc{(}ENU\textsc{)}:} ENU is the ratio of files transferred to files requested. It is an important parameter in quantifying the effectiveness of any replication strategy in data grids. ENU is measured as follows: $$\frac{N_{remote\ file\ accesses} + N_{file\ replications}}{N_{remote\ file\ accesses} + N_{local\ file\ accesses}}$$
  where:\\
- $N_{remote\ file\ accesses}$ is the number of times jobs make a remote access to get the required data for their execution. In such a case, a file is accessed remotely without replicating it locally.\\
- $N_{file\ replications}$ is the number of replicas created during jobs execution.\\
- $N_{local\ file\ accesses}$ is the number of times where the required data for a job execution is in the storage element of the site where the job is being carried out.

Note that this measure ranges from 0 to 1. Moreover, the closer its value is to 0, the more efficient is the strategy.
\end{itemize}

The obtained results are given in Table \ref{Results for the DR2 strategy and performance difference of the distribution D2 compared to the distribution D1} and Table \ref{Results for the Periodic Optimiser strategy and performance difference of the distribution D2 compared to the distribution D1}, for respectively the DR2 strategy and the Periodic Optimiser strategy.\footnote{The performance difference associated to RT for example is equal to $\displaystyle\frac{RT_{D1} \ - \ RT_{D2}}{RT_{D1}}$ $\times$ 100.}

\begin{table}[htbp]
\begin{center}
{\twlrm
        \begin{tabular}{|c||c|c|c||c|c|c|}
					\hline
									 \textbf{Number}   &   RT for \textbf{$D1$} &    RT for \textbf{$D2$}   &    {\textbf{Performance}} & {ENU for \textbf{$D1$}}   &   {ENU for \textbf{$D2$}} &     {\textbf{Performance}} \\\textbf{of jobs}
&\textsc{(}in \textit{ms}\textsc{)}&\textsc{(}in \textit{ms}\textsc{)}  &    {\textbf{difference \textsc{(}in \%\textsc{)}}}&& &     {\textbf{difference \textsc{(}in \%\textsc{)}}} \\
				 \hline \hline
				
	  {100}         &          {4 155}        &      {2 412}    &        {41.90}           &   {0.46}  &    {0.30}  &    {34.70}  \\
	       \hline
	  {500}         &            {9 345}        &      {7 009}    &          {35.70}           &    {0.36}  &    {0.25}  &   {30.50}  \\
	       \hline
	       {1 000}         &          {18 620}        &      {13 289}    &          {28.61}           &   {0.24}  &   {0.17}  &  {29.10}  \\
	       \hline
	       {1 500}         &       {24 804}        &   {19 275}    &       {22.20}           &  {0.19}  &    {0.14}  &   {26.31}  \\
\hline
	       {2 000}         &          {28 011}        &      {22 788}    &     {18.61}           &   {0.13}  &   {0.10}  &   {23.00}  \\	       \hline
	       	       {3 000}         &        {42 689}        &       {36 076}    &          {15.40}           &   {0.10}  &    {0.08}  &  {20.00}  \\	       \hline
        \end{tabular}}
				\caption{\label{Results for the DR2 strategy and performance difference of the distribution D2 compared to the distribution D1}Results for the DR2 strategy and performance difference of the distribution $D2$ compared to the distribution $D1$}
\end{center}
\end{table}

\begin{table}[htbp]
\begin{center}
{\twlrm
        \begin{tabular}{|c||c|c|c||c|c|c|}
					\hline
									 \textbf{Number}   &   RT for \textbf{$D1$} &    RT for \textbf{$D2$}   &    {\textbf{Performance}} & {ENU for \textbf{$D1$}}   &   {ENU for \textbf{$D2$}} &     {\textbf{Performance}} \\\textbf{of jobs}
&\textsc{(}in \textit{ms}\textsc{)}&\textsc{(}in \textit{ms}\textsc{)}  &    {\textbf{difference \textsc{(}in \%\textsc{)}}} && &     {\textbf{difference \textsc{(}in \%\textsc{)}}} \\
				 \hline \hline
				
	   {100}         &         {3 320}        &     {2 045}    &        {38.43}           &  {0.43}  &   {0.30}  &    {30.20}  \\
	       \hline
	 {500}         &         {7 135}        &       {5 194}    &          {27.21}           &    {0.26}  &   {0.18}  &  {30.77}  \\
	       \hline
	       {1 000}         &           {9 917}        &    {8 077}    &          {18.55}           &    {0.18}  &    {0.13}  &    {27.74}  \\

	       \hline
	       {1 500}         &         {13 455}        &     {11 360}    &        {15.57}       &   {0.13}  &   {0.10} &      {23.07}  \\
\hline
	       {2 000}         &        {16 990}        &      {14 733}    &      {13.20}           &    {0.09}  &   {0.07}  &  {22.24}  \\	       \hline
	       	       {3 000}         &          {22 109}        &       {20 844}    &         {5.70}           &    {0.06}  &   {0.05}  &    {16.60}  \\	       \hline
        \end{tabular}}
				\caption{\label{Results for the Periodic Optimiser strategy and performance difference of the distribution D2 compared to the distribution D1}Results for the Periodic Optimiser strategy and performance difference of the distribution $D2$ compared to the distribution $D1$}
\end{center}
\end{table}

The performance difference obtained for both strategies reaches 41.90\% for the RT and 34.70\% for the ENU. This is due to the starting distribution -- either $D1$ or $D2$ -- because the distribution is the only parameter that varies in these experiments. We can then clearly see that the quality of the initial distribution of replicas has an impact on the results of evaluation metrics.

\subsection{Importance of the evaluation of the distribution quality}

We now consider the general case when a distribution $D2$ is supposed better than a distribution $D1$ from the view point of replicas location. 
In such a situation, $D2$ should make easier the task of the replication strategy thereby promoting the obtained results. On the contrary, $D1$ causes difficulties to the replication strategy which degrades the evaluation results. So, if we consider the RT and ENU evaluation metrics, we will obtain $RT_{D1}$ $>$ $RT_{D2}$ and $ENU_{D1}$ $>$ $ENU_{D2}$ despite the fact that we are applying the same strategy $S$ \textsc{(}as it is the case for the obtained results in Tables \ref{Results for the DR2 strategy and performance difference of the distribution D2 compared to the distribution D1} and \ref{Results for the Periodic Optimiser strategy and performance difference of the distribution D2 compared to the distribution D1}\textsc{)}.



The study of the distribution quality makes it then possible to propose a new type of \textit{intra-strategy} evaluation unlike several other assessment methods which are \textit{inter-strategies}. This can be carried out by studying the effect of different distributions on a given strategy. Moreover, the evaluation metrics proposed in the literature are mainly dedicated to the quantitative aspect related to the grid use. In this situation, it will be interesting to offer an evaluation, according to a set of criteria, dedicated to the qualitative aspect -- like the distribution quality of replicas -- which is of paramount importance since affecting the grid performances, and hence the quantitative aspect.
It is then interesting to find answers to the following key questions:
\begin{itemize}
  \item How can we \textit{quantitatively} evaluate the quality of a placement of a given replica w.r.t. a set of criteria?
  \item How can we obtain a global view on the quality of the distribution of a set of replicas?
  \item How can we assess the effect of a replication strategy on the quality of distribution and vice-versa, \textit{i.e.}, the effect of a distribution on the performance results of a strategy?
  \item How can we offer a distribution-aware evaluation of replication strategies using commonly used metrics \textsc{(}such as RT and \textsc{ENU}\textsc{)}?\\
\end{itemize}

The following paragraphs offer a detailed description of the proposed metrics based on the aforementioned general idea as well as some representative experimental results highlighting the specificities of each metric.

In this regard, we model the files of the grid by considering that, at a given time $t$, there are $p$ distinct original files, called \textit{master files}, distributed on different sites in the grid. This set of files is then as follows: \textit{Master files} = \{$F_1$, $F_2$, ..., $F_i$, ..., $F_p$\}.

For each file $F_i$, there are $m_i$ replicas. For example, the replicas of $F_i$ are: \{$F_{i1}$, $F_{i2}$, ..., $F_{ij}$, ..., $F_{i{m_i}}$\}. The number of replicas $m_i$ varies from one file to another.\\Each replica $F_{ij}$ is requested from $n$ different sites \{$S_1$, $S_2$, ..., $S_k$, ..., $S_n$\} such that the value of $n$ is different from one replica to another.

We also take into account that the dynamicity of the grid, \textit{i.e.}, that a site can connect to or disconnect from the grid at any moment. In addition, a site can sometimes be overloaded since it is highly accessed which hampers it from satisfying the whole set of requests it receives. At the time $t$, the availability of each site is then calculated based on the data grid history using the following formula \cite{Chettaoui2014}:
\begin{equation}\label{eqn:disponibilite}
P_{S_i} = 1- \frac{\#Failure_{S_i}}{\#SiteRequest_{S_i}}
\end{equation}
where ${\#Failure}_{{S}_{i}}$ is the number of times that the site ${S}_{i}$ is failed, and ${\#SiteRequest}_{{S}_{i}}$ indicates the  total number of times that the site ${S}_{i}$ is  requested.\\The value of  ${P}_{{S}_{i}}$ increases with the increase of the availability of the site ${S}_{i}$. Indeed, a replication strategy is considered more efficient w.r.t. the distribution quality point of view if it places replicas in highly available sites.


\section{Assessment of the distribution quality based on remote accesses}

\subsection{Definition and properties of the dedicated metric}
The first metric, denoted RQD as an acronym of \textit{Remote accesses-based Quality of Distribution}, has for main purpose to assess the cost of remote accesses to replicas carried out by sites. It then advantages a strategy which places a replica in a highly available site, where the replica will be frequently requested by remote sites through high bandwidths. In this respect, the evaluation of the distribution quality will go through two steps:
\begin{itemize}
  \item A first step allows evaluating the placement of a single replica.
  \item A second step generalizes this evaluation to all replicas of the grid.
\end{itemize}
The result of the second step is then an evaluation of the distribution quality for the entire grid from the view point of remote accesses.\\

For this purpose, we define a weight associated with each replica $F_{ij}$, denoted ${RQD}_{{F}_{ij}}$. This weight reflects the placement importance of the replica in the grid. It is computed by browsing through all its $n$ remotely requesting sites. The weight is then determined as follows \cite{JNCA2015}:
\begin{equation}\label{eqn:RQD}
RQD_{F_{ij}} =  \sum_{k=1}^{n} \textsc{(} \#Request_{S_{k},F_{ij}} \times BW_{S_{k},S_{F_{ij}}} \times P_{S_{F_{ij}}}\textsc{)}
\end{equation}
where: \textit{${\#Request}_{\textit{${S}_{k}$},\textit{${F}_{ij}$}}$} denotes the \textit{number of requests} for a replica \textit{${F}_{ij}$} by a site \textit{${S}_{k}$}, \textit{${BW}_{\textit{${S}_{k}$},\textit{${S}_{\textit{${F}_{ij}$}}$}}$} represents the \textit{bandwidth} available between the site \textit{${S}_{\textit{${F}_{ij}$}}$} that contains the replica \textit{${F}_{ij}$} and a requesting site \textit{${S}_{k}$}, and \textit{${P}_{\textit{${S}_{\textit{${F}_{ij}$}}$}}$} is the \textit{availability} of the site \textit{${S}_{\textit{${F}_{ij}$}}$}.\\

In order to perform the second step -- browsing all replicas of the grid -- we have two possible alternatives  \cite{JNCA2015}:
\begin{itemize}
  \item The first is to compute the weight average of the replicas of each master file. After that, we obtain the ${RQD}_{files}$ metric which is equal to the average of the weights of the different master files.
  \item The second consists in browsing all replicas of each site to compute the weight average of the replicas per site. This results in the ${RQD}_{sites}$ metric which is equal to the average of the site weights for the different sites of the grid.
\end{itemize}
These two instances ${RQD}_{files}$ and ${RQD}_{sites}$ offer two different viewpoints. On the one hand, considering ${RQD}_{files}$, a file which is rare in the grid, \textit{i.e.} that having a low number of associated replicas, is considered more important than a file which has several replicas. Indeed, constrained by their number, the quality of the placement of the replicas associated to rare files becomes more critical for the overall grid performances. On the other hand, ${RQD}_{sites}$ promotes a site that plays an important role in the grid and sanctions the evaluation results if there are many inactive sites. In this respect, the dispersion of popular files in a large number of sites is considered as a strong point of the distribution since it allows load balancing, avoids workload peaks and improves the fault tolerance. On the contrary, grouping replicas in a limited number of sites can generate significant losses due to difficulties of accessing to these sites.\\

It should be noted that with every improvement in the distribution quality, the value of RQD increases, where RQD denotes ${RQD}_{files}$ and ${RQD}_{sites}$ when we describe their common properties. We denote ${RQD}_{before}$ the value of RQD before the use of the replication strategy and ${RQD}_{after}$ the value of RQD once the replication strategy has completed its execution. The result of taking the difference between ${RQD}_{after}$ and ${RQD}_{before}$ \textsc{(}\textit{i.e.}, ${RQD}_{after}$ - ${RQD}_{before}$\textsc{)} is  positive if the strategy improves the distribution quality, while it is negative when the strategy degrades the distribution quality.\\

To summarize, let us recall that both RQD alternatives only concentrate on remote accesses while neglecting local ones. They are mainly designed in order to quantify to which degree the replicas are well placed from the view point of remotely requesting sites.
RQD then models the bandwidth consumption while conditioning it by the availability of sites containing the remotely accessed replicas so as to sanction the strategies which rely on not always available sites. Noteworthily, both alternatives ${RQD}_{files}$ and ${RQD}_{sites}$ can be applied on all types of replication strategies. However, they are more appropriate for periodic replication strategies. Indeed, such strategies perform replications at the end of each period of time while there is no replication during the period. Hence, in each period, a strategy has a specific distribution of file replicas. RQD can then give key information on the evolution of the distribution quality from a period to another.

\subsection{Experimental evaluation of the RQD metric}

Using the OptorSim simulator, we prove experimentally the utility of the RQD metric through performing several experiments on replication strategies proposed in the literature. Further details on the used simulator as well as the simulation parameters and obtained experimental results can be found in \cite{JNCA2015}.

\begin{figure}[!t]
\parbox{15.cm}{\hspace{-0.5cm}
\includegraphics[scale=0.5]{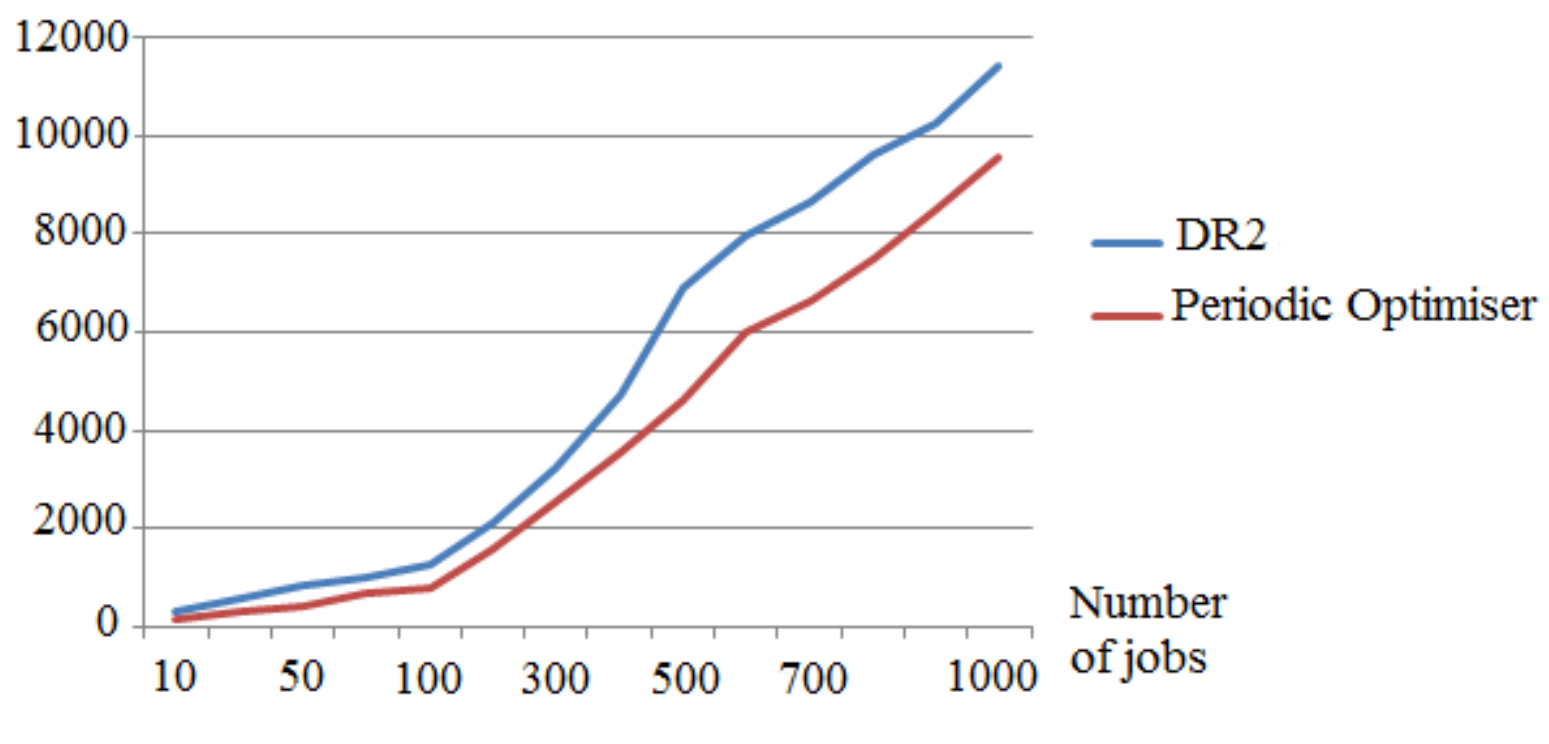}
\includegraphics[scale=0.5]{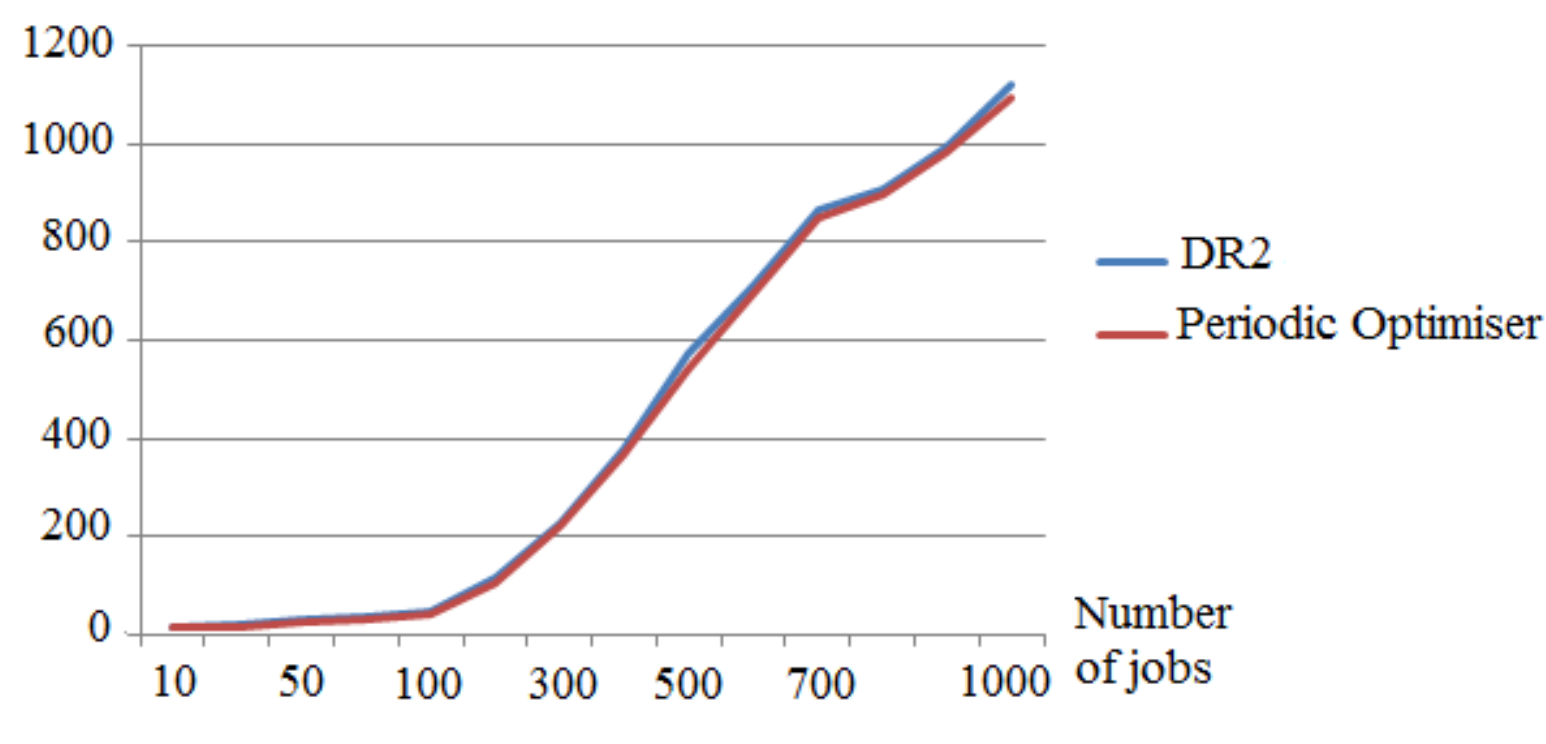}}
\caption{Results of $RQD_{files}$ \textsc{(}Left\textsc{)} and $RQD_{sites}$ \textsc{(}Right\textsc{)} for the DR2 and Periodic Optimiser strategies \label{fig_RQD}}
\end{figure}

Using the default configuration given by CMS testbed \cite{BellEtAl_vs_OptorSim}, we evaluate two strategies, namely DR2 \cite{DR2} and Periodic Optimiser \cite{PeriodicOptimiser}.
It is, however, important to note that the study applied on these two strategies can be generalized to any strategy. The evaluation of these replication strategies with ${RQD}_{\textit{files}}$ and ${RQD}_{\textit{sites}}$ gives the results shown in Table \ref{res1} and depicted in Figure \ref{fig_RQD}.


\begin{table}[h]
\begin{center}
        \begin{tabular}{|c||c|c||c|c|}
                 \hline
	      &   \multicolumn{2}{c||}{$RQD_{files}$} &   \multicolumn{2}{c|}{$RQD_{sites}$}   \\
					\hline
					\hline
					 {\textbf{Number of jobs}}   &   {\textbf{DR2}}   &   \textbf{{Periodic}  {Optimiser}} &  {\textbf{DR2}}   &   \textbf{{Periodic} {Optimiser}}  \\
				 \hline
	     {10}         &         {310.5}        &       {151.1}    &          {14.9}           &    {12.2}     \\
 \hline
	     {25}         &         {581.6}        &       {294.3}    &          {19.4}           &    {16.5}     \\
 \hline
	     {50}         &         {825.1}        &       {436.8}    &          {28.9}           &    {26.4}     \\
 \hline
	     {75}         &         {1 019.8}        &       {662.0}    &          {36.7}           &    {31.5}     \\
 \hline
	     {100}         &         {1 260.7}        &       {810.6}    &          {44.2}           &    {41.6}     \\
 \hline
	     {200}         &         {2 100.7}        &       {1 600.0}    &          {118.0}           &    {107.0}     \\
 \hline
	     {300}         &         {3 241.2}        &       {2 555.0}    &          {229.2}           &    {222.6}     \\
 \hline
	     {400}         &         {4 709.6}        &       {3 527.4}    &          {375.6}           &    {366.8}     \\
	       \hline
	    {500}         &           {6 919.5}          &    {4 639.9}     &        {571.1}            &     {543.0}   \\
	       \hline
	  {600}         &           {7 954.8}     &      {6 021.3}      &        {710.0}          &  {697.9}    \\
	       \hline
	     {700}         &         {8 675.0}        &       {6 633.8}    &          {866.0}           &    {851.4}     \\
 \hline
	     {800}         &         {9 619.3}        &       {7 480.0}    &          {910.0}           &    {898.0}     \\
 \hline
	     {900}         &         {10 237.5}        &       {8 514.4}    &          {992.6}           &    {981.2}     \\
 \hline
	     {1 000}         &         {11 428.4}        &       {9 577.3}    &          {1 120.0}           &    {1 093.6}     \\
	       \hline
        \end{tabular}
				\caption{\label{res1} Results of $RQD_{files}$ and $RQD_{sites}$ for the DR2 and Periodic Optimiser strategies}
        \end{center}
\end{table}

Table \ref{res1} shows that as far as the number of jobs increases, the values of both metrics increase. This is explained by the augmentation of the number of carried out remote accesses when there are more executed jobs. 
However, the comparison between two strategies can only be made for the same set of jobs \textsc{(}\textit{i.e.}, for a same line in Table \ref{res1}\textsc{)}.\\

Let us also note that for the same number of jobs and for the same strategy, the value of $RQD_{files}$ is largely greater than that of $RQD_{sites}$, reaching in average more than 17 times for DR2 and 12 times for Periodic Optimiser. Indeed, $RQD_{sites}$ evaluates each site independently of the other sites to give an importance to a site that plays an important role in the grid through the replicas it contains. At the same time, it sanctions the evaluation results if there are many inactive sites.\\

If we examine the results of both strategies, we find that the DR2 strategy improves the quality of the distribution more than the Periodic Optimiser strategy. However, the gap in the obtained results of both strategies is greater when we use ${RQD}_{files}$, although decreasing from 51.33\% to 16.19\% with the increase of the number of jobs. The use of ${RQD}_{sites}$ also promotes DR2 too, although with less difference between both tested strategies \textsc{(}decreasing from 18.12\% to 2.35\% when the number of jobs reaches 1 000\textsc{)}.\\

The obtained results can be argued by the following two complementary facts:
\begin{itemize}
  \item The DR2 strategy gives more importance to rare files compared to Periodic Optimiser \textsc{(}\textit{cf.} columns 2 and 3 of Table \ref{res1}\textsc{)}. Indeed, if we analyze the DR2 strategy in the third phase of its first step \cite{DR2}, we find that DR2 replicates a file only if it considers it \textit{rare} compared to the other files. On the contrary, when looking at the Periodic Optimiser strategy, we find that this strategy does not take this criterion into account. This disadvantages Periodic Optimiser compared to DR2 strategy when performing the evaluation using ${RQD}_{files}$.
  \item The DR2 strategy gives poor dispersion of files against Periodic Optimiser \textsc{(}\textit{cf.} columns 4 and 5 of Table \ref{res1}\textsc{)}. Indeed, once the DR2 strategy decides to replicate a file \textsc{(}\textit{i.e.}, in its second step\textsc{)}, it seeks to minimize the number of hops between sites. On its side, Periodic Optimiser does not impose any condition that may affect the dispersion of the files in the grid, which promotes it when considering the dispersion of replicas. This explains the reduced difference between DR2 and Periodic Optimiser when evaluating with ${RQD}_{sites}$.\\
\end{itemize}

In addition to remote accesses, the next metrics we propose take into consideration the efficiency of replicas placements w.r.t. local accesses. Indeed, a given replica may be highly requested by the holding site while less remotely required. This should also be considered as a positive aspect w.r.t. the distribution quality.

\section{Assessment of the replications effect on the distribution}
\subsection{Definition and properties of the dedicated metric}
We now propose a different way for evaluating a given replication strategy by instantly assessing its effects on the distribution quality. The proposed evaluation is based on the fact that a strategy should maximize high-quality replications within strategic sites as much as possible and, hence, avoid costly remote accesses. For this purpose, the following two complementary considerations are taken into account.

On the one hand, a replication operation is considered beneficial from the view point of the distribution quality if it creates a replica of a file $F_k$ in a highly available site $S_i$ which frequently requests for $F_k$ through a low bandwidth connecting it with the holding site $S_j$. This hence avoids the high cost of accessing this replica remotely and offers future efficient local accesses.

On the other side, the strategy should minimize costly remote accesses. Note that a remote access operation to a file $F_k$ is considered costly if it is frequently carried out by a site $S_i$ towards a holding site $S_j$ of a low availability and through a low bandwidth.\\

In the general case, repeated remote accesses are more costly than replications and cause an increasing consumption of bandwidth. So, whenever the cost of remote accesses increases, performed replications must compensate the losses the remote accesses have caused. Otherwise, the strategy proves to be inefficient in its influence on the distribution quality. In fact, maximizing the performed number of replications allows avoiding potential future remote accesses and offers more local ones which make data access more efficient. The actual effect of a replication of the distribution is however conditioned by the right choice of the files to replicate as well as the associated locations of new replicas.\\

We call the dedicated metric RED as an acronym of \textit{Replications Effect on the Distribution} in reference to the effect of the replications performed by the strategy on the quality of replicas distribution in the grid. The associated formula for RED is the  following \cite{JNCA2015}:
\begin{equation}\label{eqn:RED}
RED = \frac{\sum_{}^{Replication}  \frac{\#Request_{S_i,F_k} \times P_{S_i}}{BW_{S_i,S_j}}}{\sum_{}^{Replication}  \frac{\#Request_{S_i,F_k} \times P_{S_i}}{BW_{S_i,S_j}} + \sum_{}^{RemoteAccess} \frac{\#Request_{S_i,F_k}}{BW_{S_i,S_j} \times P_{S_j}}}
\end{equation}
where:\\
- \textit{Replication}: the set of all the performed replication operations.\\
- \textit{RemoteAccess}: the set of all the remote access operations.\\

In this respect, the worst strategy, from the RED metric point of view, is the one that makes no replications since it does not generate any gain to the distribution quality. RED is then equal to 0. While, the considered best strategy is the one that
replicates each remotely requested file in the requesting site. In this case, there is no remote access and RED is then equal to 1. Hence, RED lies between 0 and 1 while the higher is its value the better is the distribution quality.\\

RED can give an evaluation at any instant, which allows to easily follow the evolution of the distribution quality over time. RED can then be applied to all types of replication strategies while being more suitable for non-periodic replication strategies. Indeed, on the one hand, RED evaluates instantly the distribution quality. On the other hand, such strategies perform either a replication or a remote access each time a site makes a request about not locally available data. Hence, RED allows evaluating the effect of such a request immediately. It is however important to mention that although it offers a consideration for the induced local accesses through performed replications compared to the RQD metric which only concentrates on remote accesses, RED is relatively computationally more expensive.

\subsection{Experimental evaluation of the RED metric}

\begin{figure}[htbp]
\centering
\includegraphics[scale=0.5]{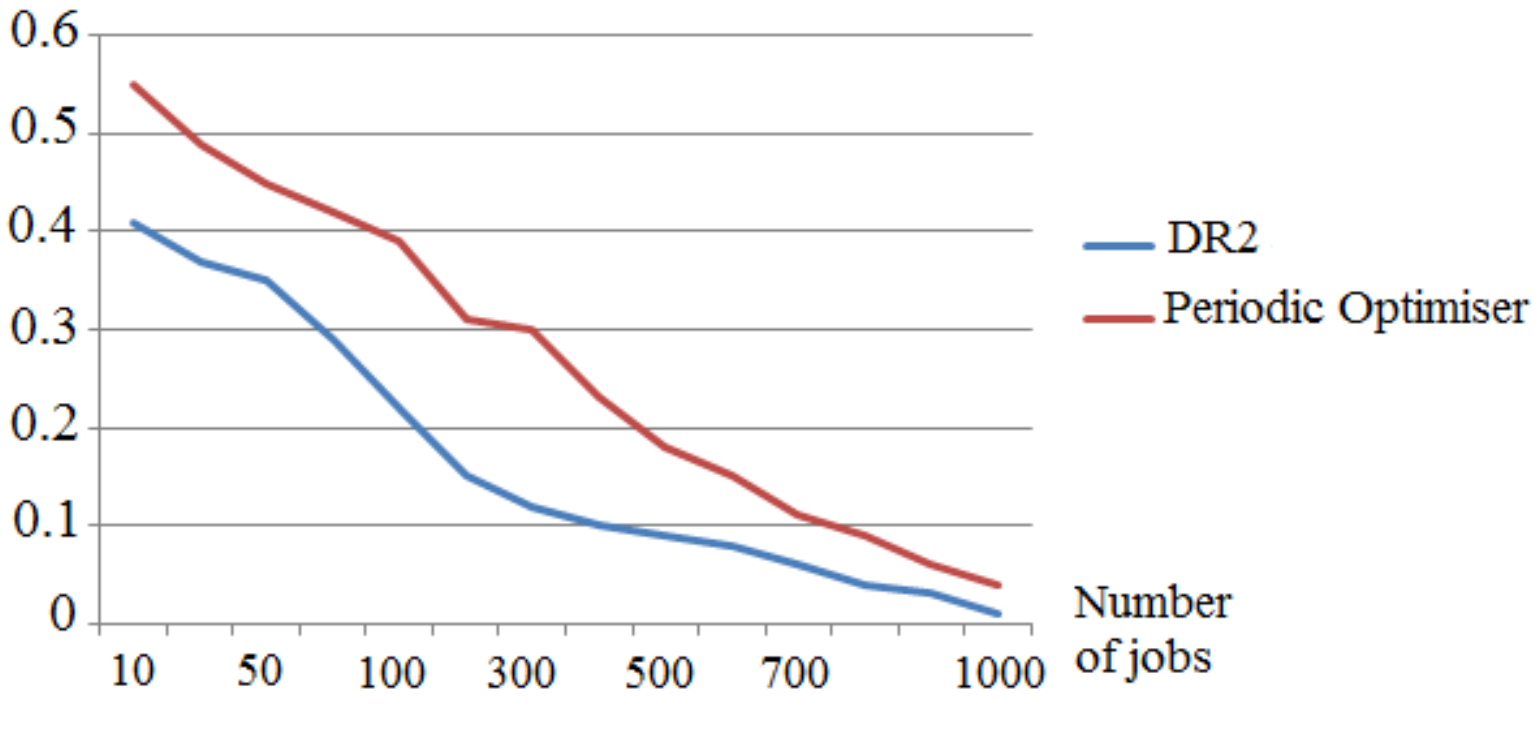}
\caption{\label{hisRed} Results of RED for the DR2 and Periodic Optimiser strategies}
\end{figure}

The evaluation results of the RED metric for the replication strategies DR2, Periodic Optimiser, Least Recently Used \textsc{(}LRU\textsc{)} and No Replication under the CMS testbed configuration are presented in Table \ref{ev} and Figure \ref{hisRed}.

\begin{table}[h]
\begin{center}
        \begin{tabular}{|c||c||c|c||c|}
           \hline
	        \textbf{{Number} {of jobs}}   &   \textbf{{No} {Replication}}   &    \textbf{{DR2}} &   \textbf{{Periodic} {Optimiser}} &   \textbf{LRU}   \\
	       \hline
	       \hline
	       {10}       &    {0}               &  {0.41}     &    {0.55}   &    { 1}\\
 \hline
	       {25}       &    {0}               &  {0.37}     &    {0.49}   &    {1}\\
 \hline
	       {50}       &    {0}               &  {0.35}     &    {0.45}   &    {1}\\
 \hline
	       {75}       &    {0}               &  {0.29}     &    {0.42}   &    {1}\\
 \hline
	       {100}       &    {0}               &  {0.22}     &    {0.39}   &    {1}\\
 \hline
	       {200}       &    {0}               &  {0.15}     &    {0.31}   &    {1}\\
 \hline
	       {300}       &    {0}               &  {0.12}     &    {0.30}   &    {1}\\
	       \hline
	       {400}       &      {0}             &   {0.10}   &   {0.23}   &    {1}   \\
	       \hline
	       {500}        &       {0}             &   {0.09}   &    {0.18}   &    {1} \\
 \hline
	       {600}       &    {0}               &  {0.08}     &    {0.15}   &    {1}\\
 \hline
	       {700}       &    {0}               &  {0.06}     &    {0.11}   &    {1}\\
 \hline
	       {800}       &    {0}               &  {0.04}     &    {0.09}   &    {1}\\
 \hline
	       {900}       &    {0}               &  {0.03}     &    {0.06}   &    {1}\\
 \hline
	       {1 000}       &    {0}               &  {0.01}     &    {0.04}   &    { 1}\\
	       \hline
        \end{tabular}
        		\caption{\label{ev} Evaluation of some replication strategies with RED}
        \end{center}
\end{table}

According to the obtained results, it is worth noting that both strategies No Replication and LRU respectively represent both typical cases for which we obtain the minimum and the maximum bounds of the RED values. Indeed, the first strategy does not perform replications and as a consequence RED is equal to 0, while the second always replicates and thus the value of RED is 1.

It should also be noted that the influence of Periodic Optimiser and DR2 on the distribution quality turns out great in the beginning. Then, progressively, it begins to decrease. This is explained by the fact that the replication strategy makes a lot of replication at the beginning because it is processing new file requests. In addition, carried out replications are those allowing to avoid costly remote accesses.
Then, progressively, once the number of jobs increases, the strategy will meet almost the same file requests without having to perform a lot of replications as at the beginning. 
This offers at the start a high value of RED which decreases over time since the best replicas are already created.\\

A comparison between the results of Table \ref{ev} and those of Table \ref{res1} \textsc{(}\textit{cf.} page \pageref{res1}\textsc{)} allows to note an important point: with the RQD evaluation, the DR2 strategy was better. However, with the RED evaluation, the Periodic Optimiser strategy has become better with a gain varying from 22.22\% to 75.00\%.\\Knowing that the RQD metric ignores local accesses while RED considers such local accesses induced by carried out replications, we can deduce that replications made by Periodic Optimiser provide local accesses more than those of DR2. This important point is proved experimentally in \cite{JNCA2015} through new RED-oriented variants of these strategies as well as several other experiments.

\section{Assessment of the distribution quality based on remote and local accesses}
\subsection{Definition and properties of the dedicated metric}

The general idea behind this supplementary assessment method is to firstly quantify the total cost of remote accesses in the grid. Then, the evaluation of the distribution quality according to this cost is performed while taking into consideration the performed local accesses. Such accesses reflect the positive aspect of the distribution. Indeed, a local access avoids the use of bandwidth and consequently reduces the response time. As a consequence, we consider it as a desirable thing that increases the value of the distribution quality metric. On the contrary, a remote access causes losses in several aspects such as the bandwidth consumption, the response time and the data availability. So, we consider it as an undesirable thing that should be avoided as much as possible. However, we must distinguish remote accesses because some of them are very costly while others are less expensive.\\

Based on this idea, a new metric called DisQ, as an acronym of \textit{Distribution Quality}, is introduced whose proposal goes through two steps: \textsc{(}$i$\textsc{)} evaluating the placement of a single replica, \textsc{(}$ii$\textsc{)} generalizing this evaluation to all replicas of the grid and, subsequently, offering an evaluation of the distribution quality for the entire grid.

We then propose the following weight for a given replica ${F}_{ij}$ which is placed in a site $S_{F_{ij}}$ \cite{IJWET2016}:
\begin{equation}
ReplicaWeight_{F_{ij}} = \frac{TotalLocalCost_{F_{ij}}}{TotalLocalCost_{F_{ij}} + TotalRemoteCost_{F_{ij}}}
\label{+1}
\end{equation}
where TotalLocalCost$_{F_{ij}}$ quantifies the total cost of accessing the replica $F_{ij}$ locally, \textit{i.e.}, from the holding site $S_{F_{ij}}$. While TotalRemoteCost$_{F_{ij}}$ indicates the total cost of accessing $F_{ij}$ remotely, \textit{i.e.}, from its requesting sites.\\

A high value of ${ReplicaWeight}_{{F}_{ij}}$ represents the goodness of the placement of the replica ${F}_{ij}$. The cost of a remote access operation is computed in the same manner as for the RED metric \textsc{(}\textit{cf.} Equation \ref{eqn:RED}, page \pageref{eqn:RED}\textsc{)}. While TotalLocalCost$_{F_{ij}}$ is equal to its number of local requests. Indeed, for local access operations, the bandwidth is not considered while the availability of $S_{F_{ij}}$ is considered equal to 1.\\
For each master file, its weight is equal to the average of its replicas weights. Then, the DisQ metric allowing the evaluation of the distribution of all grid replicas is equal to the average of the resulting weights of the grid master files.\\

It is important to note that for a given replica $F_{ij}$, 0 $<$ $ReplicaWeight_{F_{ij}}$ $\leq$ 1. Since we take the average to obtain the file weight, this latter is also encompassed between 0 and 1. The same applies for the value of DisQ. In addition, the higher the value of DisQ is, the better is the distribution quality. Moreover, comparing for a given strategy the value of DisQ at two instants $t1$ and $t2$ \textsc{(}with $t1$ < $t2$\textsc{)} allows to guess its effect on the distribution \textsc{(}\textit{i.e.}, whether it improves the distribution or not during its invocation between $t1$ and $t2$\textsc{)}.

\subsection{Experimental evaluation of the DisQ metric}


Quantifying the quality of the distribution using DisQ proves the impact of the distribution on the results of a replication strategy. In this respect, using four different distributions, Table \ref{timePO} and Table \ref{ENUPO} show the values in \textit{ms} of the response time \textsc{(}RT\textsc{)} and those of the ENU metric for Periodic Optimiser. The experiments were carried out with a low-quality distribution \textsc{(}DisQ = 0.2\textsc{)}, a high-quality distribution \textsc{(}DisQ = 0.8\textsc{)}, and two other distributions \textsc{(}DisQ = 0.4 and DisQ = 0.6\textsc{)}. The same work is done in Table \ref{timeDR2} and Table \ref{ENUDR2} for the DR2 strategy. Note that further details on the carried out experiments related to DisQ can be found in \cite{CoopIS2016,IJWET2016}.

\begin{table}[htbp]
     \begin{center}
        \begin{tabular}{|p{3cm}||p{2.2cm}|p{2.2cm}|p{2.2cm}|p{2.2cm}|}
                         \hline

            \makebox[\linewidth][c]{\textbf{Number of jobs}}  &     \makebox[\linewidth][c]{\textbf{DisQ = 0.2}} &     \makebox[\linewidth][c]{\textbf{DisQ = 0.4}} &   \makebox[\linewidth][c]{\textbf{DisQ = 0.6}}    & \makebox[\linewidth][c]{\textbf{DisQ = 0.8}} \\
 \hline             \hline

  \makebox[\linewidth][c]{100} & \makebox[\linewidth][c]{3 314} & \makebox[\linewidth][c]{2 840} & \makebox[\linewidth][c]{2 495}  &   \makebox[\linewidth][c]{2 148} \\
   \hline


 \makebox[\linewidth][c]{500} & \makebox[\linewidth][c]{7 679} &  \makebox[\linewidth][c]{6 807} & \makebox[\linewidth][c]{5 915}  &  \makebox[\linewidth][c]{5 452}\\

  \hline


 \makebox[\linewidth][c]{1 000} & \makebox[\linewidth][c]{10 576} &  \makebox[\linewidth][c]{9 742} & \makebox[\linewidth][c]{8 880}  &  \makebox[\linewidth][c]{8 184}\\
   \hline

 \makebox[\linewidth][c]{1 500} & \makebox[\linewidth][c]{14 089} &  \makebox[\linewidth][c]{13 511} & \makebox[\linewidth][c]{11 836}  &  \makebox[\linewidth][c]{10 735}\\

  \hline
 \makebox[\linewidth][c]{2 000} & \makebox[\linewidth][c]{17 172} &  \makebox[\linewidth][c]{16 504} & \makebox[\linewidth][c]{15 369}  &  \makebox[\linewidth][c]{14 930}\\

\hline

        \end{tabular}
				\caption{\label{timePO}Response time values obtained in \textit{ms} by the Periodic Optimiser strategy using different distributions}
	 \end{center}
\end{table}

\begin{table}[htbp]
     \begin{center}
        \begin{tabular}{|p{3cm}||p{2.2cm}|p{2.2cm}|p{2.2cm}|p{2.2cm}|}
                 \hline

            \makebox[\linewidth][c]{\textbf{Number of jobs}}  &     \makebox[\linewidth][c]{\textbf{DisQ = 0.2}} &     \makebox[\linewidth][c]{\textbf{DisQ = 0.4}} &   \makebox[\linewidth][c]{\textbf{DisQ = 0.6}}    & \makebox[\linewidth][c]{\textbf{DisQ = 0.8}} \\
 \hline             \hline

  \makebox[\linewidth][c]{100} & \makebox[\linewidth][c]{0.43} & \makebox[\linewidth][c]{0.39} & \makebox[\linewidth][c]{0.34}  & \makebox[\linewidth][c]{0.29} \\
   \hline


 \makebox[\linewidth][c]{500} & \makebox[\linewidth][c]{0.26} &  \makebox[\linewidth][c]{0.24} & \makebox[\linewidth][c]{0.21}  &  \makebox[\linewidth][c]{0.18}\\

  \hline


 \makebox[\linewidth][c]{1 000} & \makebox[\linewidth][c]{0.18} &  \makebox[\linewidth][c]{0.17} & \makebox[\linewidth][c]{0.15}  &  \makebox[\linewidth][c]{0.13}\\

  \hline
     \makebox[\linewidth][c]{1 500} & \makebox[\linewidth][c]{0.15} & \makebox[\linewidth][c]{0.13} & \makebox[\linewidth][c]{0.12}  & \makebox[\linewidth][c]{0.11} \\
   \hline
 \makebox[\linewidth][c]{2 000} & \makebox[\linewidth][c]{0.09} &  \makebox[\linewidth][c]{0.08} & \makebox[\linewidth][c]{0.07}  &  \makebox[\linewidth][c]{0.07}\\

\hline
        \end{tabular}
				\caption{\label{ENUPO}ENU values obtained by the Periodic Optimiser strategy using different distributions}
	 \end{center}
\end{table}

%


\begin{table}[!t]
     \begin{center}
        \begin{tabular}{|p{3cm}||p{2.2cm}|p{2.2cm}|p{2.2cm}|p{2.2cm}|}
                         \hline

             \makebox[\linewidth][c]{\textbf{Number of jobs}}  &     \makebox[\linewidth][c]{\textbf{DisQ = 0.2}} &     \makebox[\linewidth][c]{\textbf{DisQ = 0.4}} &   \makebox[\linewidth][c]{\textbf{DisQ = 0.6}}    & \makebox[\linewidth][c]{\textbf{DisQ = 0.8}} \\
 \hline             \hline

  \makebox[\linewidth][c]{100} & \makebox[\linewidth][c]{4 203} & \makebox[\linewidth][c]{3 615} & \makebox[\linewidth][c]{2 984}  & \makebox[\linewidth][c]{2 554} \\
   \hline


 \makebox[\linewidth][c]{500} & \makebox[\linewidth][c]{9 416} &  \makebox[\linewidth][c]{8 304} & \makebox[\linewidth][c]{7 721}  &  \makebox[\linewidth][c]{7 056}\\
  \hline


 \makebox[\linewidth][c]{1 000} & \makebox[\linewidth][c]{17 584} &  \makebox[\linewidth][c]{16 092} & \makebox[\linewidth][c]{14 314}  &  \makebox[\linewidth][c]{12 801}\\

  \hline
     \makebox[\linewidth][c]{1 500} & \makebox[\linewidth][c]{22 115} & \makebox[\linewidth][c]{20 284} & \makebox[\linewidth][c]{18 360}  & \makebox[\linewidth][c]{16 492} \\
   \hline
 \makebox[\linewidth][c]{2 000} & \makebox[\linewidth][c]{27 391} &  \makebox[\linewidth][c]{25 367} & \makebox[\linewidth][c]{23 610}  &  \makebox[\linewidth][c]{21 426}\\

\hline

        \end{tabular}
				\caption{\label{timeDR2}Response time values obtained in \textit{ms} by the DR2 strategy using different distributions}
	 \end{center}
\end{table}

\begin{table}[!t]
     \begin{center}
        \begin{tabular}{|p{3cm}||p{2.2cm}|p{2.2cm}|p{2.2cm}|p{2.2cm}|}
                        \hline
            \makebox[\linewidth][c]{\textbf{Number of jobs}}  &     \makebox[\linewidth][c]{\textbf{DisQ = 0.2}} &     \makebox[\linewidth][c]{\textbf{DisQ = 0.4}} &   \makebox[\linewidth][c]{\textbf{DisQ = 0.6}}    & \makebox[\linewidth][c]{\textbf{DisQ = 0.8}} \\
 \hline             \hline

  \makebox[\linewidth][c]{100} & \makebox[\linewidth][c]{0.48} & \makebox[\linewidth][c]{0.42} & \makebox[\linewidth][c]{0.35}  & \makebox[\linewidth][c]{0.30} \\
     \hline


 \makebox[\linewidth][c]{500} & \makebox[\linewidth][c]{0.36} &  \makebox[\linewidth][c]{0.33} & \makebox[\linewidth][c]{0.29}  &  \makebox[\linewidth][c]{0.24}\\
  \hline


 \makebox[\linewidth][c]{1 000} & \makebox[\linewidth][c]{0.24} &  \makebox[\linewidth][c]{0.22} & \makebox[\linewidth][c]{0.20}  &  \makebox[\linewidth][c]{0.17}\\

  \hline
     \makebox[\linewidth][c]{1 500} & \makebox[\linewidth][c]{0.19} & \makebox[\linewidth][c]{0.17} & \makebox[\linewidth][c]{0.15}  & \makebox[\linewidth][c]{0.13} \\
   \hline
 \makebox[\linewidth][c]{2 000} & \makebox[\linewidth][c]{0.13} &  \makebox[\linewidth][c]{0.11} & \makebox[\linewidth][c]{0.10}  &  \makebox[\linewidth][c]{0.09}\\

\hline

        \end{tabular}
				\caption{\label{ENUDR2}ENU values obtained by the DR2 strategy using different distributions}
	 \end{center}
\end{table}


In the general case, the lower the value of the distribution quality is, the worse are the evaluation results of replication strategies. On the contrary, the better is the distribution quality, the more interesting are the results. This illustrates the influence of the initial distribution, pre-existing before invoking the strategy, on the results of the evaluation metrics.\\


The next section is dedicated to the proposal of an evaluation process of replication strategies taking into account the quality of replicas distribution already existing when a strategy is invoked.

\section{Distribution quality-aware evaluation of replication strategies}

\subsection{Correction process and adaptation of evaluation metrics}


The proposed metrics allow evaluating the quality of the replicas distribution at each given instant. We then exploit this key feature towards an objective evaluation of each replication strategy \cite{IJWET2016}. This is carried out by taking into account the influence of the initial distribution that has been found, before launching the strategy, on the evaluation results of this latter through metrics like RT, ENU, etc.

\begin{figure}[htbp]
\centering
\includegraphics[scale=0.335]{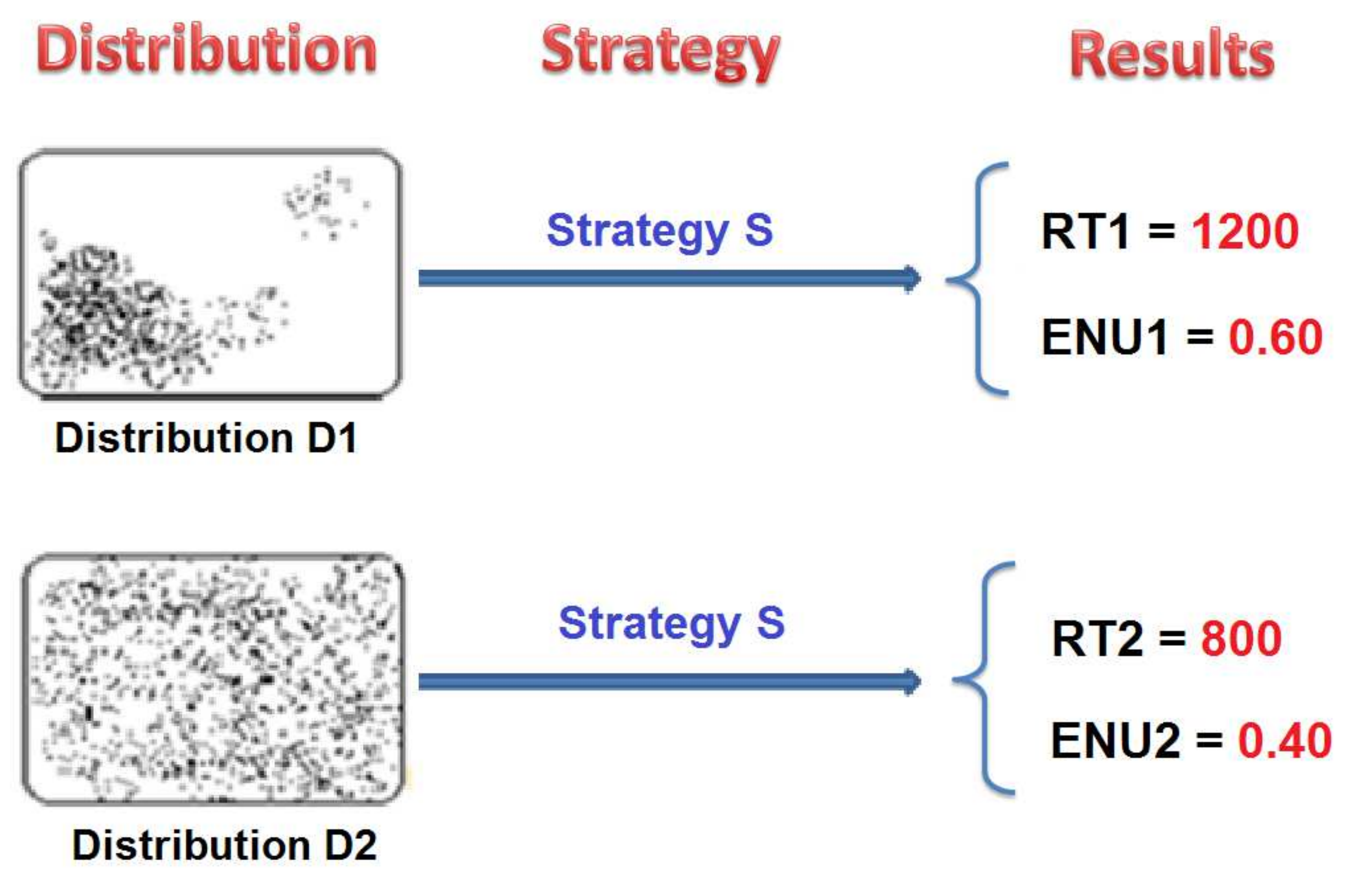}
\caption{Evaluation results without considering the initial distribution}\label{other1}
\end{figure}
\begin{figure}[htbp]
\centering
\includegraphics[scale=0.335]{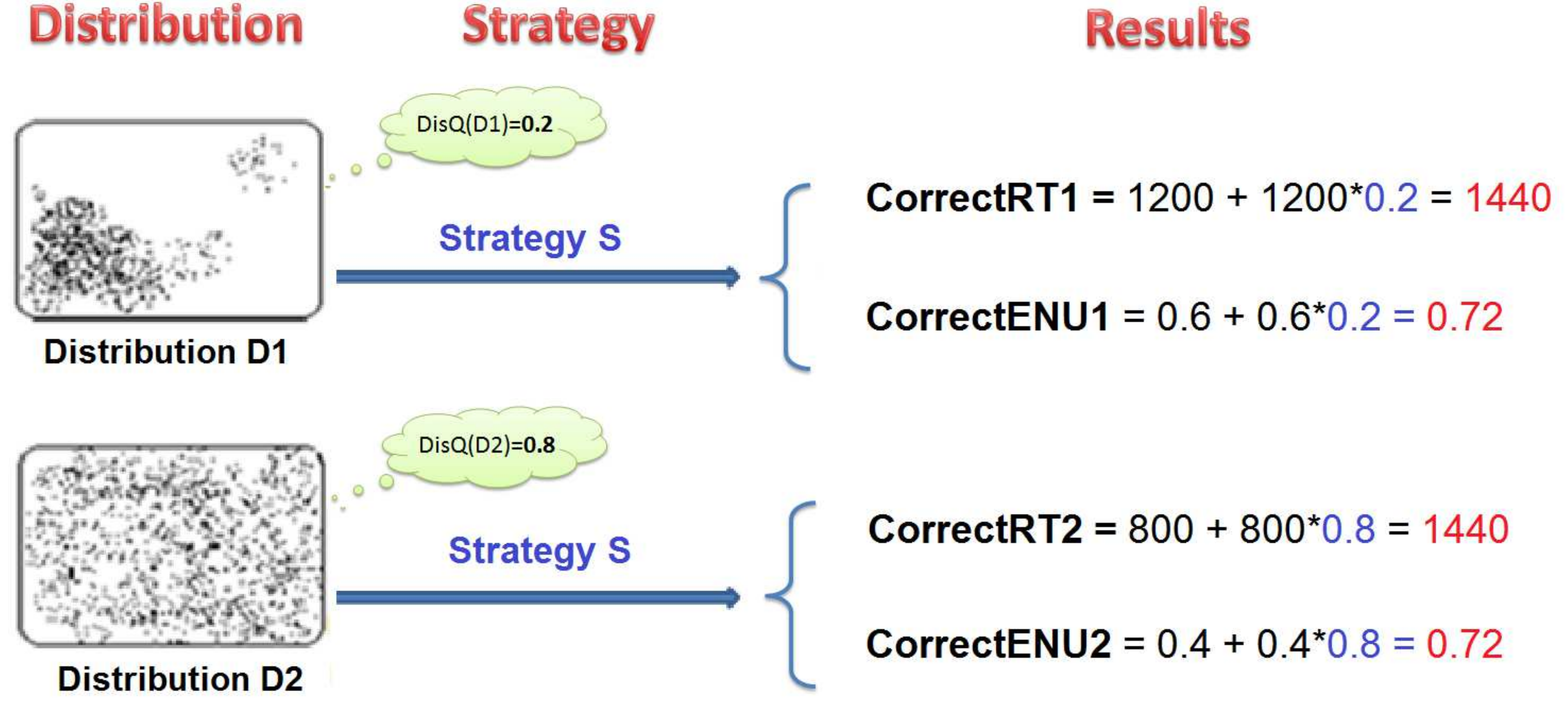}
\caption{Evaluation results when considering the initial distribution}\label{other2}
\end{figure}

In order to illustrate the general idea, we use the DisQ metric. Let us in this respect take the illustrative example in Figure \ref{other1}. The application of the same strategy $S$ with two different distributions $D1$ and $D2$ gives two different results for the RT and the ENU metrics. However, the obtained experimental values do not necessarily reflect the real performances of the strategy. In our opinion, we should consider the effect of the distribution quality initially found to ensure the objectivity of the evaluation. In this sense, let us consider the quality values of both distributions $D1$ and $D2$ found before launching the strategy as equal to 0.2 and 0.8, respectively. We then attempt to correct the evaluation results according to the quality of both distributions as depicted in Figure \ref{other2}. In this way, the results can be adjusted by making them independent from the initial distribution.\\

In the general case, to correct the value of a given evaluation metric, \textit{i.e.}, to obtain the correct metric value, we need two information: \textsc{(}$i$\textsc{)} the value of $DisQ$ before starting the strategy, and, \textsc{(}$ii$\textsc{)} the initial value, denoted $Metric$, of the metric itself, \textit{i.e.}, that assessed while not considering the quality of the distribution on which the strategy was executed.

The main idea of the correction consists of excluding the influence of the initial distribution from the obtained result. In the instance of the correction process we propose here, the new metric value that we consider more appropriate since it is quality-aware is calculated as follows:
\begin{equation}
CorrectMetric = Metric - DisQ \times Metric \times X
\label{correction}
\end{equation}
where the value $X$ $\in$ \{-1, 1\} defined by:\\
$\left\{
\begin{array}{l}
 X = 1 \ \ \mbox{if the metric adopts a direct aspect where a high value of the metric is a good}\\\mbox{sign for the strategy.}\\
 X = -1 \ \ \mbox{if the metric adopts a counterproductive aspect where a high value of the me-}\\\mbox{tric is a bad sign for the strategy.}
\end{array}
\right.$\\\\




To illustrate the idea, we create two instances of this formula for both evaluation metrics, namely the RT and the ENU, as follows:
\begin{equation}\label{eqn:CoorectRT}
CorrectRT = RT - DisQ \times RT \times \textsc{(}-1\textsc{)} = RT + DisQ \times RT
\end{equation}
\begin{equation}\label{eqn:CorrectENU}
CorrectENU = ENU - DisQ \times ENU \times \textsc{(}-1\textsc{)} = ENU + DisQ \times ENU
\end{equation}

\bigskip

\subsection{Experimental evaluation}

In order to show the utility of the correction process, we observe the values, obtained after 100 jobs, taken by the RT and ENU metrics starting from two different distributions: the quality of the first is equal to 0.2 while that of the second is equal to 0.8. The experiments are performed using the DR2 and Periodic Optimiser strategies.

Table \ref{Simulations with different distributions} shows the obtained results of when varying the initial distribution quality. The evaluation results show the superiority of DR2 over Periodic Optimiser. This is due to the quality of the initial distribution which is in the favor of DR2 since equal to 0.8, while it is hampering the performances of Periodic Optimiser since equal to 0.2. Indeed, if we compare the obtained results when the same distribution quality is used \textsc{(}of a value either equal to 0.2, 0.8, or any other value\textsc{)} through focusing on Table \ref{timePO} and Table \ref{timeDR2} \textsc{(}\textit{cf.} page \pageref{timePO}\textsc{)}, the superiority of Periodic Optimiser over DR2 is shown for the different number of jobs.

\begin{table}[!t]
     \begin{center}
        \begin{tabular}{|p{2cm}|p{3cm}|p{2.5cm}|p{2.5cm}|}
                         \hline
            \makebox[\linewidth][c]{\textbf{DisQ}}  &     \makebox[\linewidth][c]{\textbf{Strategy}} &     \makebox[\linewidth][c]{\textbf{RT} \textsc{(}in \textit{ms}\textsc{)}} &   \makebox[\linewidth][c]{\textbf{ENU}}    \\
 \hline             \hline
\makebox[\linewidth][c]{0.2} & \makebox[\linewidth][c]{Periodic Optimiser} &  \makebox[\linewidth][c]{3 314} &  \makebox[\linewidth][c]{0.43} \\
\hline
\makebox[\linewidth][c]{0.8} &  \makebox[\linewidth][c]{DR2} &  \makebox[\linewidth][c]{2 554} &  \makebox[\linewidth][c]{0.30} \\
\hline
        \end{tabular}
				\caption{\label{Simulations with different distributions} Simulations with different distributions}
	 \end{center}
\end{table}

However, we do not always have the opportunity to compare strategies under the same distribution quality. 
We need as a consequence to correct the results using DisQ in order to obtain a more reliable evaluation. Table \ref{corrected} illustrates the results once corrected. The corrected values prove the superiority of Periodic Optimiser strategy. The reliability of the results is then ensured without having to start with the same distribution.

\begin{table}[htbp]
     \begin{center}
        \begin{tabular}{|p{1cm}|p{2.9cm}|p{2.2cm}|p{3.1cm}|p{1.2cm}|p{1.8cm}|}
                         \hline
            \makebox[\linewidth][c]{\textbf{DisQ}}  &     \makebox[\linewidth][c]{\textbf{Strategy}} &     \makebox[\linewidth][c]{\textbf{RT} \textsc{(}in \textit{ms}\textsc{)}} & \makebox[\linewidth][c]{\textbf{CorrectRT} \textsc{(}in \textit{ms}\textsc{)}} &   \makebox[\linewidth][c]{\textbf{ENU}}  &          \makebox[\linewidth][c]{\textbf{CorrectENU}}  \\
 \hline \hline
 \makebox[\linewidth][c]{0.2} & \makebox[\linewidth][c]{Periodic Optimiser} &  \makebox[\linewidth][c]{3, 314} & \makebox[\linewidth][c]{3, 976} &  \makebox[\linewidth][c]{0.43} & \makebox[\linewidth][c]{0.51}\\
\hline
\makebox[\linewidth][c]{0.8} &  \makebox[\linewidth][c]{DR2} &  \makebox[\linewidth][c]{2, 554}  & \makebox[\linewidth][c]{4, 597} &  \makebox[\linewidth][c]{0.30} & \makebox[\linewidth][c]{0.54} \\
\hline
        \end{tabular}
				\caption{\label{corrected} Corrected results}
	 \end{center}
\end{table}

Moreover, several obtained experimental results highlighted in \cite{IJWET2016} show the accuracy of the correction process. Here, for the sake of brevity, we only present as a sample the results obtained for the RT metric and for the Periodic Optimiser strategy. These  experiments aim to highlight the need to exclude the influence of the distribution on the evaluation results so we can find convergent results even with different distributions.

\begin{figure}[htbp]
\centering
\includegraphics[scale=0.7]{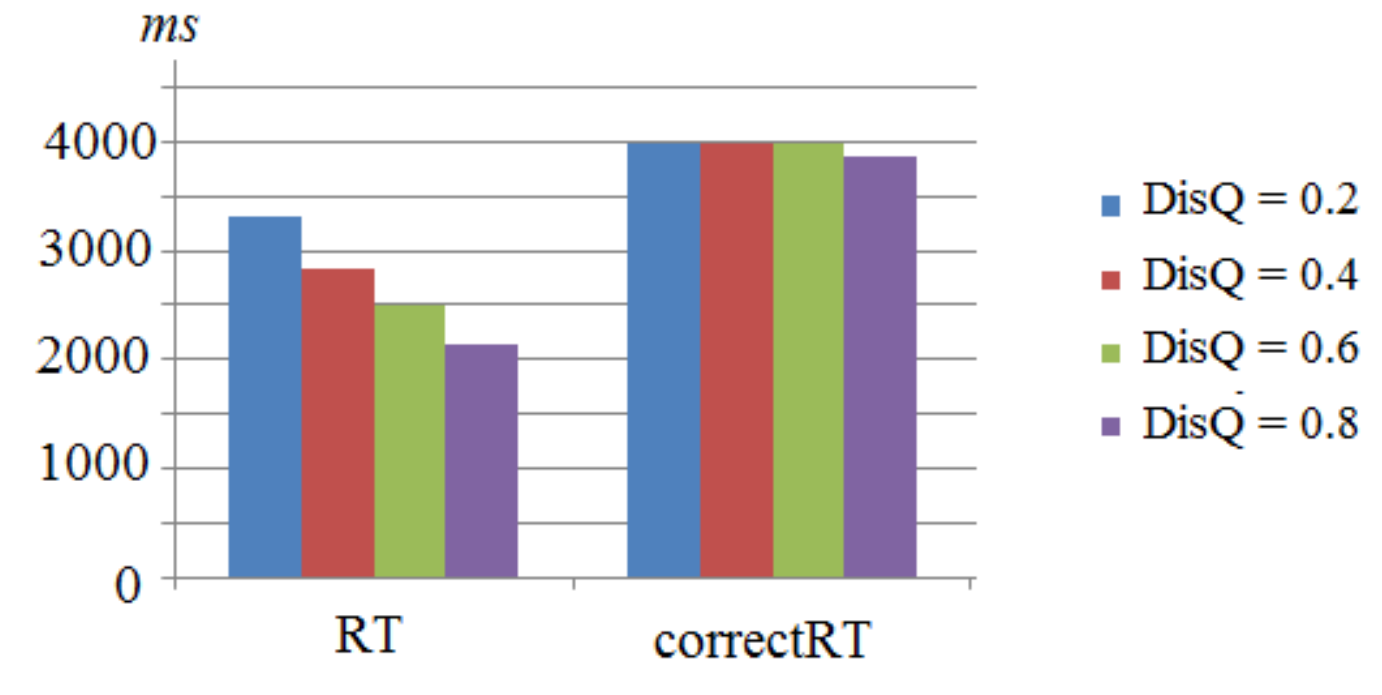}
\vspace{-0.7cm}
\caption{Difference between the obtained RT values before and after the correction for Periodic Optimiser}\label{DisQ_RT_PO}
\end{figure}

We set the number of jobs to 100 and we vary the initial distribution quality used. The results obtained through the RT metric, on the one hand, and the corrected variant that takes into consideration the distribution quality value, on the other hand, are then compared. Table \ref{responsecorrectionPO} and Figure \ref{DisQ_RT_PO} depict the obtained results.\\
Although they are taken from the same strategy and after the execution of the same jobs, the difference between the RT values before the correction is large and reaches 54.28\%. It is known that this difference is due to the variation of the distribution that the strategy works on. On the other side, the difference between the RT values once the correction process applied becomes negligible and is at most equal to 3.26\%.\\

\begin{table}[htbp]
     \begin{center}
        \begin{tabular}{|p{2.5cm}||p{3.5cm}|p{3.5cm}|}
                         \hline

  \makebox[\linewidth][c]{\textbf{DisQ}} & \makebox[\linewidth][c]{\textbf{RT} \textsc{(}in \textit{ms}\textsc{)}} & \makebox[\linewidth][c]{\textbf{CorrectRT} \textsc{(}in \textit{ms}\textsc{)}}  \\
\hline \hline
  \makebox[\linewidth][c]{0.2} & \makebox[\linewidth][c]{3 314} & \makebox[\linewidth][c]{3 976} \\
\hline
  \makebox[\linewidth][c]{0.4} & \makebox[\linewidth][c]{2 840} & \makebox[\linewidth][c]{3 976} \\
\hline
 \makebox[\linewidth][c]{0.6} &  \makebox[\linewidth][c]{2 495} & \makebox[\linewidth][c]{3 992} \\
\hline
  \makebox[\linewidth][c]{0.8} & \makebox[\linewidth][c]{2 148} & \makebox[\linewidth][c]{3 866} \\
\hline
        \end{tabular}
				\caption{\label{responsecorrectionPO} RT and CorrectRT values in \textit{ms} obtained for the Periodic Optimiser strategy using different distributions}
	 \end{center}
\end{table}

\section{Conclusion}

In this chapter, we mainly addressed the key problem related to the impact of the replicas distribution within sites on the results of replication strategies. 
We then proposed a new criterion having a direct impact on the results of quantitative evaluation metrics, namely the quality of the replicas distribution on the grid.
We then quantified the distribution quality by proposing new metrics dedicated to this purpose. In this respect, we evaluated the effect of a given replication strategy on the distribution quality through different ways: \textsc{(}$i$\textsc{)} the RQD metric mainly based on remote accesses while taking into consideration the consumed bandwidth as well as the availability of sites, 
\textsc{(}$ii$\textsc{)} the RED metric instantly assessing the effect of the decisions taken by a strategy on the distribution quality w.r.t. the induced efficiency of data access, \textsc{(}$iii$\textsc{)} the DisQ metric taking into consideration the costs of already performed remote and local accesses. The effect of a given distribution on the performance of replication strategies was also studied in-depth. A paramount contribution of this chapter consisted then in the proposal of a new distribution-aware correction process of the results of the evaluation metrics of replication strategies. Extended experiments were carried out using the OptorSim simulator in order to validate our theoretical results.


\chapter{Data Mining-based Replication Strategies}\label{chapter_quality_of_distribution}
\setcounter{footnote}{0}

\section{Introduction and motivations}

Replication in data grids has attracted a great deal of attention of many
researchers. Many strategies have then been proposed in the literature and several surveys have been conducted \cite{surveyFGCS2012,surveyreplication2013,Grace,MoHa2015.1}. However, most of existing replication strategies are based on single file granularity and neglect correlation among different data files.

Actually, in many applications, data files may be correlated in terms of accesses and have to be considered together in order to reduce the access cost \cite{TuCorrelated2014}.
Indeed, the analysis of data usage in several real data grids \cite{fileCule2008,Ko2007} revealed the existence of strong correlations between files, \textit{i.e.}, jobs tend to simultaneously request a set of correlated files.
Therefore, replicating or prefetching correlated files in data grids, can help in minimizing bandwidth consumption, communication latency, response time as well as
resource consumption \cite{Allcock2002}. Taking into account knowledge about data correlations in replication strategies can then help on achieving both efficient data management and performance. Indeed, this can be achieved if correlated data, \textit{i.e.}, data that are processed frequently together, are stored or replicated at the same or nearby locations \cite{Iritani2012}. In fact, complex queries of jobs or users typically access jointly to a set of data which can require access multiple sites to retrieve them all. For example, Kayyoor \textit{et al.} have proposed a workload-aware replica selection and replica placement algorithms that attempt to minimize the total resources consumed in a distributed environment through co-locating correlated data items \cite{KayyoorSWORD}. It has also been recognized in various contexts, like file systems \cite{conf/hpdc/XiaFJTW08}, distributed systems \cite{EventCorrelationsInLogs2012,WuJPDC2013}, cloud systems \cite{6522257,Ye:2013:GGB:2480362.2480438}, as well as web environment \cite{TuCorrelated2014,Zhuo2003927}, that taking into account knowledge about data correlations in replication and caching strategies can help in achieving both efficient data management and performance.\\

The knowledge about data correlation can be extracted from historical data using techniques of the data mining field. Data mining techniques have proved to be a powerful tool facilitating the extraction of meaningful knowledge from large data sets \cite{Han:2011:DMC:1972541,zaki2014dataminingbook}. In the literature, only few works have used data mining techniques to explore file correlations in data grids. The main approaches for mining file correlations can be classified into two categories: frequent sequence mining and association rule mining. Unfortunately, these two approaches were not effectively used. Indeed, works based on frequent sequence mining cannot fully reveal semantic file correlations. In fact, they focus on access file sequences without considering their contexts, \textit{i.e.}, the semantic aspect of file usage, namely a file is requested by which job. In addition, they do not distinguish the different requests issued from the different grid jobs and consider them as successive file accesses. On the other side, various studies have shown the limits of association rule mining based on the support and confidence approach.\\

We deem that the choice of the data mining technique results in three main objectives to be satisfied: extracting knowledge with high quality, as rapidly as possible, while improving the performance of the replication process. However, these objectives are often contradictory. It is therefore of paramount importance to find the best trade-off between the quality of the extracted knowledge and the performance of the strategy. In order to overcome these drawbacks, we propose to use correlated pattern mining to explore file correlations. The main advantage of this approach is to strengthen the support measure with a more powerful measure that reveals whether the items in a set are really correlated or not, namely the \textit{all-confidence} measure \cite{Omie03}.
This measure enables generating highly correlated patterns while reducing significantly the number of patterns mined and offering opportunities for an efficient mining of patterns compared to other correlation measures \cite{pakdd2013,Omie03,Xiong06hypercliquepattern}. In addition, we only exploit \textit{maximal} \textit{frequent} correlated patterns and not all correlated patterns.\\

The design of this strategy is at the crossroads of several contributions we have done. On the one hand, through several theoretical results consolidated by experiments, the utility of mining reduced sets \textit{aka} concise representations \cite{IDA2012} of frequent \cite{DS2010,TSI2012} and rare \cite{PAKDD2012,IDA2015} correlated patterns is shown. On the other hand, in \cite{EAAI2016}, a complete guideline offers a roadmap for the application of data mining techniques in data grid replication. The proposed guideline fully describes all steps of the process including: \textsc{(}$i$\textsc{)} the translation from the data grid context to the data mining one and vice-versa, \textsc{(}$ii$\textsc{)} the choices to be taken into account and the constraints stemming from the data grid and the data mining contexts. It also clarifies the main choices to be made when designing a data mining based replication strategies covering the following key steps:
\begin{itemize}
  \item First step: the grid data selection and preprocessing.
  \item Second step: the data mining step.
  \item Third step: the replication step based on the results of the previous step.\\
\end{itemize}

We then propose a new dynamic periodic decentralized data replication strategy, called RSCP \cite{JSS2015}.\footnote{RSCP is the acronym of \textit{Replication Strategy based on Correlated Patterns}.} RSCP follows the key steps of the guideline and considers a set of correlated files as granularity for replication. To overcome the limits of the existing replication strategies based on data mining techniques, the proposed strategy:
\begin{itemize}
\item constructs a binary context based on history of file access job, from which the mining process is performed,
\item relies on a new algorithm for mining maximal frequent correlated pattern mining in order to infer grid file correlations. Each maximal frequent correlated pattern represents a maximal set of files frequently appearing simultaneously and whose correlation degree exceeds a dedicated minimum threshold of the used correlation measure. This allows a given file to belong to distinct mined patterns since it can be simultaneously correlated to different sets of files.
\item considers both file associated parameters \textsc{(}\textit{e.g.} size\textsc{)} and grid topology parameters \textsc{(}\textit{e.g.} bandwidth\textsc{)}.
\end{itemize}

The evaluation metrics we analyze in the experiments are mean job execution time, effective network usage, total number of replications, hit ratio and percentage of storage filled. Using the OptorSim simulator, extensive experimentations show that our proposed strategy has better performance in comparison to other strategies under most of access patterns.

\section{State of the art of strategies based on data mining techniques}
Several studies proposing replication strategies taking into account file correlations have been conducted. In our work, we mainly focus on those based on data mining techniques, which are succinctly presented in ascending order by year of publication. We refer readers to \cite{Han:2011:DMC:1972541,zaki2014dataminingbook} for a detailed study on data mining techniques. Interested readers are also referred to our work proposed in \cite{ICCS2015,EAAI2016} for further details on these strategies and on some other replication strategies taking into account file correlations and adopting other techniques than those of data mining.

In this respect, it is important to note that in addition to the criteria mentioned in the previous chapter offering a classification of replication strategies \textsc{(}\textit{cf.} page \pageref{criteriaForReplicationStrategiesClassification}\textsc{)}, those strategies relying on data mining techniques can further be classified according to the following specificities of the data mining approach on which they are based:
\begin{itemize}
\item \textbf{Adopted data mining technique:} a given strategy can exploit several kinds of knowledge, extracted through a data mining technique in order to perform a given task, like looking for association between file attributes, clustering grid sites into disjoint groups, etc.

\item \textbf{Data used in the data mining process:} a given data mining process uses as input data extracted from the data grid on which the strategy is applied.

\item \textbf{Explicit knowledge:} this consists in the extracted patterns from the data mining algorithm, like frequent sequences, association rules, clusters, etc.

\item \textbf{Data mining periodicity:} the data mining algorithm is triggered
    at each file request or at each period.

\item \textbf{Centralized/Decentralized data mining:} the data mining algorithm can be launched by each grid site independently of the others. This represents the decentralized case. In the centralized case, only a central site has the key role of running the data mining part.
\end{itemize}

Among the strategies which use data mining techniques we briefly describe the main following ones.
The main idea of the PRA strategy \cite{Preftching08} is to make use of the characteristics that members in a virtual organization have similar interests in files to carry out a better replication optimization. The algorithm is described as following: when a site $S_{i}$ does not have a file locally, it requests a remote site $S_{j}$. This latter receives the request and transfer the file to the former site. At the same time, it finds the adjacent files of the requested file by applying frequent pattern sequence mining technique on the file access sequence data base. At last, a message containing the list of adjacent files will be sent to the site $S_{i}$ that will choose adjacent files to replicate. 

The RSCA strategy \cite{RSCA09} is based on the existence of correlations among the
data files accessed according to the access history of
grid users. At the first stage, a clustering analysis is conducted on the file access history of sites in the grid over a period of time. The outputs of this operation are
correlated file sets related to the access habits of users. At the second stage, a replication is done on the basis of those sets, which achieves the aim of prefetching and buffering data. The clustering method adopted is used to group into equivalence classes all the files that are similar according to a given equivalence relation. The set of files in the same equivalence class are called correlative file sets. 


The ARRA strategy \cite{ARRA10} is introduced in two parts. In the first part, access behaviors of data-intensive jobs are analyzed based on the Apriori algorithm \cite{Agra94}. In the second part, replica replacement rules are generated and applied. Accessed data files are considered as items of the database mined through Apriori, while each transaction is composed by the required data files of each data-intensive job. Noteworthily, in \cite{7026895}, a more recent strategy based on Apriori for association rule mining has been proposed.

By considering spatial locality, PHFS \cite{PHFS11} uses predictive techniques to predict the future usage of files and then pre-replicates them in hierarchical manner on a path from the source to the client in order to increase locality in access. File correlations are inferred from previous access patterns by association rules and clustering techniques of data mining. PHFS operates in three steps. In the first, file access information are collected in the root site. In the second step, data mining techniques are applied on log files. Finally, whenever a client requests a file, PHFS
finds the predicted subsequent requests after this request.

The main idea of PDDRA \cite{PDDRA12} is the same of the PRA strategy. Based on file access history, PDDRA predicts future needs of grid sites and prefetches a sequence of files to the requester grid site. As a consequence, the next time that this site needs a file, it will be locally available. PDDRA consists of three phases: storing file access patterns, requesting a file, and finally performing replication, prefetching and replacement.

The major idea of the BSCA strategy proposed in \cite{BSCA_13} is to prefetch frequently
accessed files and their associated files to the location near the access site. It finds out the correlation between the data files through data access number and data access serial. In addition of the use of the access numbers, BSCA is also based on the support and the confidence measures used for mining association rules. This strategy has two sub algorithms: data mining algorithm and replication algorithm. Once the data mining algorithm is applied to identify frequent files, support and confidence of association rules between these frequent files are computed. If the support and the confidence values between files exceed respective minimum thresholds, frequent files and their associated ones are replicated. 

\section{Analysis of the state of the art data mining-based strategies}
The criteria used for classifying the surveyed replication strategies based on data mining techniques are as follows:
\begin{itemize}
  \item[$\bullet$] Periodicity,
  \item[$\bullet$] Centralized/Decentralized replication,
  \item[$\bullet$] Adopted data mining technique,
  \item[$\bullet$] Data mining periodicity,
  \item[$\bullet$] Centralized or decentralized data mining,
  \item[$\bullet$] Data used in the data mining process,
  \item[$\bullet$] Main parameters used in the replication strategy.
\end{itemize}
Table \ref{tabComp} summarizes the properties of the different strategies. In this table, \textit{C} stands for \textit{Centralized}, \textit{D} for \textit{Decentralized}, \textit{DM} for \textit{Data Mining}, \textit{ARM} for \textit{Association Rule Mining}, and \textit{FSM} for \textit{Frequent Sequence Mining}.\\

\begin{table}[htbp]
\parbox{12.cm}{\hspace{-0.8cm}
\twlrm
\begin{tabular}{|p{2.cm}|| p{1.6cm}| p{1cm}|| p{1.3cm}| p{1.6cm}| p{1.2cm}|p{1.7cm}||p{2.5cm}|}
	\hline
	 \makebox[\linewidth][c]{\bf Strategy}&  \makebox[\linewidth][c]{\bf Periodicity}&
\makebox[\linewidth][c]{\bf Type of} \makebox[\linewidth][c]{\bf decision} \makebox[\linewidth][c]{\bf making} &
 \makebox[\linewidth][c]{\bf DM} \makebox[\linewidth][c]{\bf technique}&
 \makebox[\linewidth][c]{\bf DM} \makebox[\linewidth][c]{\bf periodicity}&
 \makebox[\linewidth][c]{\bf C/D DM}&
\makebox[\linewidth][c]{\bf Data used} \makebox[\linewidth][c]{\bf in the DM} \makebox[\linewidth][c]{\bf process}&
 \makebox[\linewidth][c]{\bf Parameters} \\

	\hline \hline

\makebox[\linewidth][c]{\it PRA \textsc{(}2008\textsc{)}}    &\makebox[\linewidth][c]{non periodic} &     \multicolumn{1}{c||}{D} &\makebox[\linewidth][c]{FSM}  &\makebox[\linewidth][c]{non periodic} & \multicolumn{1}{c|}{D} & \makebox[\linewidth][c]{sites/accessed} \makebox[\linewidth][c]{files}  & \makebox[\linewidth][c]{file request number,}  \makebox[\linewidth][c]{frequency, confidence}   \\ \hline

\makebox[\linewidth][c]{\it RSCA \textsc{(}2009\textsc{)}}   & \makebox[\linewidth][c]{periodic}& \multicolumn{1}{c||}{C}
&\makebox[\linewidth][c]{clustering}&\makebox[\linewidth][c]{periodic}& \multicolumn{1}{c|}{C}& \makebox[\linewidth][c]{sites/accessed} \makebox[\linewidth][c]{files} &\makebox[\linewidth][c]{file request number,} \makebox[\linewidth][c]{frequency}    \\ \hline

\makebox[\linewidth][c]{\it ARRA \textsc{(}2010\textsc{)}}  & \makebox[\linewidth][c]{non periodic}   & \multicolumn{1}{c||}{D} &
\makebox[\linewidth][c]{ARM}&\makebox[\linewidth][c]{at the} \makebox[\linewidth][c]{beginning of} \makebox[\linewidth][c]{the simulation}   & \multicolumn{1}{c|}{C}&  \makebox[\linewidth][c]{jobs/accessed} \makebox[\linewidth][c]{files}&
  \makebox[\linewidth][c]{file request number,} \makebox[\linewidth][c]{frequency, confidence}
   \\ \hline

\makebox[\linewidth][c]{\it PHFS \textsc{(}2011\textsc{)}}   &\makebox[\linewidth][c]{periodic +} \makebox[\linewidth][c]{non
 periodic} & \multicolumn{1}{c||}{D} & \makebox[\linewidth][c]{ARM,} \makebox[\linewidth][c]{clustering}& \makebox[\linewidth][c]{periodic}& \multicolumn{1}{c|}{C} & \makebox[\linewidth][c]{sites/accessed} \makebox[\linewidth][c]{files}&
 \makebox[\linewidth][c]{file request number,}  \makebox[\linewidth][c]{confidence} \\ \hline

\makebox[\linewidth][c]{\it PDDRA \textsc{(}2012\textsc{)}}
& \makebox[\linewidth][c]{non periodic} & \multicolumn{1}{c||}{D}    &\makebox[\linewidth][c]{tree mining} & \makebox[\linewidth][c]{non periodic}& \multicolumn{1}{c|}{D}&  \makebox[\linewidth][c]{sites/accessed} \makebox[\linewidth][c]{files}&
 \makebox[\linewidth][c]{file size, bandwidth,}  \makebox[\linewidth][c]{file request number,}  \makebox[\linewidth][c]{frequency}    \\ \hline

\makebox[\linewidth][c]{\it BSCA  \textsc{(}2013\textsc{)}} &
\makebox[\linewidth][c]{periodic}& \multicolumn{1}{c||}{D}  & \makebox[\linewidth][c]{ARM}& \makebox[\linewidth][c]{periodic}& \multicolumn{1}{c|}{C}&\makebox[\linewidth][c]{sites/accessed} \makebox[\linewidth][c]{files}&
 \makebox[\linewidth][c]{file request number,}  \makebox[\linewidth][c]{frequency, confidence}              \\ \hline
	   \end{tabular}} \caption{Comparison of replication strategies based on data mining techniques} \label{tabComp}
\end{table}

For the sake of brevity, we only focus on the following main observations that emerge from Table \ref{tabComp}. Further details can be found in our work \cite{JSS2015}.

A first observation concerns the data used to infer file correlations. We note that most strategies consider the file access pattern of sites. In this situation, the database to be mined is composed by sites in lines and accessed files in columns. Actually, a history of file access by jobs is in our opinion a more reasonable choice to infer semantic relationships between files accessed by jobs which are executed on each site.

Another observation is associated to the parameters taken into account by the strategies. We note that there are two types of parameters. On the one hand, we find file associated parameters like number of requests and file size. On the other hand,
grid topology parameters like bandwidth are used. In this respect, several surveyed strategies mainly take into consideration parameters related to file characteristics, while rare are those which also consider grid parameters.

A key observation is that the association rule mining is the most used technique for exploring file correlations.
In these strategies, two measures of association rule quality assessment are used, namely the support and the confidence measures. However, various studies have shown the limits of association rule mining based on these quality measures \cite{IDA2015,Brin97,AMAI2010,kim2011efficient,pakdd2013}.
Indeed, the resulting set of association rules has an excessively large size, with a majority of the mined rules either redundant or not reflecting the true correlation relationship among data \cite{CheungESWA2012,comine_Lee}. 

On another side, in the case of non-periodic strategies, the data mining algorithm is triggered for each file request which would lead to negative impact on replication strategy performance, especially on the response time. Some strategies also suppose that the correlation between files is predetermined and details are lacking on the mining task of patterns to be used in the replication process. Some others are trying to address the problem of files correlation without any clue on the placement problem of the identified groups of correlated files. In addition, the assumption which consists in considering that a given file belongs only to one correlated files group is not always realistic. Indeed, multiple correlations between files are common since a given file could be correlated to several clusters of files. Each cluster may characterize the common data required by a given set of jobs.

\section{Background of correlated pattern mining}

Deeming that the file correlations extraction is a crucial step in our proposed replication strategy and that file correlations should be well evaluated, extracted patterns should be of high quality. For this reason, we adopt the correlated pattern mining approach. Correlated pattern mining was shown in the literature to be more complex and more informative than frequent pattern mining and association rules \cite{IDA2015,Cao201431,SegondB12a}.

\subsection{Preliminary concepts}
In the next points, we describe the preliminary concepts used in the data mining step of our approach which is dedicated to correlated pattern mining. We also indicate how each data grid concept is used to suit the data mining step.
 \begin{itemize}
\item \textbf{\textit{Item}}: an item is a binary attribute which indicates the presence or absence of an attribute in a specific transaction \textsc{(}\textit{aka} object\textsc{)}. In our approach, it is a logical value \textsc{(}0 or 1\textsc{)} that determines whether or not the file is requested by a specific job beyond a given minimum threshold.

\item \textbf{\textit{Transaction}}: let $\mathcal{I} = \{i_{1}, i_{2},$\ldots$ i_{n}\}$ be a set of $n$ items. A transaction contains a subset of the items in $\mathcal{I}$. It has an associated unique identifier. In our case, each transaction is a set of files required by a given job.

\item \textbf{\textit{Extraction context \textsc{(}\textit{aka} transaction database or simply database\textsc{)}}}: an extraction context in data mining is a triplet $\mathcal{K} = \textsc{(}\mathcal{O}, \mathcal{I}, \mathcal{R}\textsc{)}$, where $\mathcal{O}$ is the set of objects \textsc{(}transactions\textsc{)}, $\mathcal{I}$ is the set of attributes \textsc{(}items\textsc{)} and $\mathcal{R}$ is a binary \textsc{(}incidence\textsc{)} relation \textsc{(}\textit{i.e.}, $\mathcal{R}\subseteq \mathcal{O} \times \mathcal{I}$\textsc{)}. Each couple $\textsc{(}o,i\textsc{)}\in \mathcal{R}$ expresses that the object $o \in \mathcal{O}$ contains the item $i \in \mathcal{I}$. In the proposed strategy, accessed files are considered as items and the required files of each job constitute each transaction.

\item \textbf{\textit{Pattern}}: is a set of items. It is also commonly called \textit{itemset}.

\item \textbf{\textit{Frequent pattern}}: each itemset has an associated measure of statistical significance called \textit{support}. For an itemset $X$ $\subseteq$ $\mathcal{I}$, \textit{Supp}\textsc{(}$X$\textsc{)} is the fraction of the transactions in the extraction context containing $X$, \textit{i.e.}, the number of transactions containing $X$ divided by the number of all transactions. If an itemset has a support above the user specified minimum support, it is said to be \textit{frequent}. Otherwise it is said \textit{rare} or \textit{infrequent}. In our case, a frequent pattern represents a set of files that appear simultaneously in the file access history with a support exceeding a user specified minimum support.

\item \textbf{\textit{Correlation measure}}: A variety of correlation measures have been proposed in the literature such as \textit{lift} and \textit{$\chi^2$} \cite{Brin97}, \textit{all-confidence}, \textit{any-confidence} and \textit{bond} \cite{Omie03}, \textit{coherence} \cite{comine_Lee}, and \textit{cosine} \cite{Cao201431,re-examiniation2010} to quote but a few.

    In our work, we choose the all-confidence correlation measure. This measure initially proposed in \cite{Omie03} has the same property as the \textit{h-confidence} measure \cite{Xiong06hypercliquepattern}. It was successfully used in several real-life applications based on correlated patterns. In this respect, experimental results on some real data sets show that it is a promising correlation measure since it enables an efficient mining of highly correlated patterns while reducing significantly the number of patterns mined. Indeed, the key reason for its popular adoption is the suitable properties that it owns \cite{pakdd2013}. In this respect, several correlated pattern mining processes have been proposed based on the all-confidence correlation measure \cite{EfficientMiningFrequentlyCorrelated2012,pakdd2013,comine_Lee,Omie03,Xiong06hypercliquepattern}.

The all-confidence value of an itemset $X$ $\subseteq$ $\mathcal{I}$ is computed as follows:
\begin{center}
\textit{all-confidence}\textsc{(}$X$\textsc{)} =
$\frac{\displaystyle\textit{Supp}\textsc{(}X\textsc{)}}{\displaystyle \textit{max} \{
\textit{Supp}\textsc{(}i\textsc{)}| \ i\in X\}}$
\end{center}
This measure is one of the few interestingness measures that simultaneously hold the anti-monotony, cross-support and null-invariant properties which are described as follows:
\begin{itemize}
\item \textbf{\textit{Anti-monotone constraint}} \cite{Agra94}: a constraint $Q$ is anti-monotone if \mbox{ } $\forall  I\subseteq \mathcal{I}$, $I_1 \subseteq I$: $I$ verifies $Q \Rightarrow I_1$ verifies $Q$.

\item \textbf{\textit{Cross-support property}} \cite{Xiong06hypercliquepattern}: given a threshold  $t$ $\in$ $]0, 1[$, an itemset $I$ fulfills the cross-support property w.r.t. $t$ if $I$ contains two items $x$ and $y$ such that $\frac{\displaystyle \textit{Supp}\textsc{(}x\textsc{)}}{\displaystyle \textit{Supp}\textsc{(}y\textsc{)}} < t$.

    Both aforementioned properties offer opportunities for pruning the search space since all non-empty subsets of a correlated pattern must also be correlated. In addition, the supersets of a not correlated pattern are necessarily not correlated and thus they are not mined. This makes the mining process more efficient.

\item \textbf{\textit{Null-invariant property}} \cite{re-examiniation2010}:
A null-invariant measure allows quantifying the degree of mutual
relationships between items in a group without taking into account the items outside the group in question \cite{EfficientMiningFrequentlyCorrelated2012,kim2011efficient}. Indeed, a null-invariant correlation measure of a pattern $I$ does not depend on the number of transactions which do not contain $I$, called null transactions. This is a crucial property of a correlation measure. Indeed, it avoids the influence of null transactions on the associated values.
\end{itemize}

\item \textbf{\textit{Correlated pattern}}: a pattern is said \textit{correlated} with respect to a given correlation measure, if the value of its correlation measure is larger than or equal to a minimum threshold of correlation associated with the measure.

\item \textbf{\textit{Frequent correlated pattern}}: given an extraction context, a minimum support threshold \textit{minsupp} and a minimum threshold of correlation measure \textit{mincorr}, a pattern $X$ is said to be a \textit{frequent correlated} pattern if \textit{Supp}\textsc{(}$X$\textsc{)} $\geq$ $minsupp$ and \textit{Corr}\textsc{(}$X$\textsc{)} $\geq$ $mincorr$. In our approach, such a pattern represents a set of files whose correlation measure exceeds a given threshold and the number of simultaneous occurrences in the file access history exceeds the pre-defined minimum threshold. Both used measures -- support and all-confidence -- have then a complementary role in limiting the mined patterns to those highly correlated while frequently appearing in the extraction context.

\item \textbf{\textit{Maximal frequent correlated pattern}}: a frequent correlated pattern is said to be \textit{maximal} if it has no frequent correlated pattern as a superset.
\end{itemize}

\subsection{Maximal frequent correlated pattern mining algorithm}\label{sec:MFCP}
Although frequent correlated pattern mining discloses the correlation relationships among data and reduces significantly the number of extracted patterns, it still generates quite a large number of patterns. In this work, we rely on maximal frequent correlated pattern mining in order to reduce the number of the correlated patterns produced, while retaining the greatest sets w.r.t. set inclusion containing highly correlated files. This allows to face a problem met by traditional algorithms for mining file correlations -- either based on access sequence mining or association rule mining -- namely the generation of a huge number of patterns, many of them being redundant.\\

A new generate-and-test algorithm, called \textsc{MFCPM},\footnote{\textsc{MFCPM} is the acronym of \textit{Maximal Frequent Correlated Pattern Miner}.} allowing the extraction of maximal frequent correlated patterns is introduced. The inputs of the algorithm are: a binary context, the minimum support threshold denoted \textit{minsupp}, and the minimum correlation threshold denoted \textit{min-all-confidence}. We summarize in Table \ref{notation} the notations used in this algorithm. The pseudo-code of \textsc{MFCPM} is depicted by Algorithm \ref{IR}. The sixth instruction of \textsc{MFCPM} invokes the \textsc{Generate$\_$Next$\_\mathcal{FCP}$} procedure whose pseudo-code is
depicted by Algorithm \ref{NextMFC}. This procedure exploits both the cross-support \cite{Xiong06hypercliquepattern} and the anti-monotony \cite{Omie03} properties
of the all-confidence measure in order to efficiently discover frequent correlated patterns.\\The theoretical complexity of the \textsc{MFCPM} algorithm is in $O\textsc{(}2^n\textsc{)}$ where $n$ represents the size of the longest maximal frequent correlated pattern.

\begin{table}[!t]
\centering
 \small{
\begin{tabular}{|p{1.4cm} p{0.cm} p{9cm}|}
\hline
\textbf{Notation} & &\textbf{Description} \\ \hline 	\hline
 $X_{k}$&& A pattern $X$ of size $k$.\\
 $X$.\textit{supp}&& The support of a pattern $X$.\\
$X$.\textit{almax}&& The maximum item support of all the items in $X$.\\
$\mathcal{C}_{k}$&&  The set of candidate patterns of size $k$.\\
 $\mathcal{FCP}_{k}$& & The set of frequent correlated patterns of size $k$.\\
$\mathcal{MFCP}$&& The set of maximal frequent correlated patterns.\\
 \hline
\end{tabular}}\caption{Notations used in the \textsc{MFCPM} algorithm}\label{notation}
\end{table}

\linesnumbered
\begin{algorithm}[htbp]
 \footnotesize{
\KwData{A binary context, \textit{minsupp} and \textit{min-all-confidence}.}
\KwResult{The complete set of maximal frequent correlated patterns $\mathcal{MFCP}$.}
\Begin{
$k$ := $1$\;
$\mathcal{FCP}_{1} := \{i\in\mathcal{I} \ | \ \textit{Supp}\textsc{(}i\textsc{)}\geq
	\textit{minsupp}\}$\;
$\mathcal{MFCP}$ := $\mathcal{FCP}_{1}$\;
\While{$\mathcal{FCP}_{k} \neq \emptyset$}{\label{InRes1}
$\mathcal{FCP}_{k+1}$ := \textsc{Generate$\_$Next$\_\mathcal{FCP}$}
\textsc{(}$\mathcal{FCP}_{k}$, \textit{minsupp},
$\textit{min-all-confidence}$\textsc{)}\;
\ForEach{\textsc{(}$X_{k+1} \in \mathcal{FCP}_{k+1}$\textsc{)}}
	 {
\If{\textsc{(}$\exists X_{k} \subset X_{k+1}$ $|$ \textsc{(}$X_{k} \in \mathcal{MFCP}$ \textsc{\textsc{)}\textsc{)}} }
{remove $X_{k}$ from $\mathcal{MFCP}$}
}
 $\mathcal{MFCP}$ := $\mathcal{MFCP} \cup
\mathcal{FCP}_{k+1}$\; $k$ := $k$ +
$1$\;
}
\Return{$\mathcal{MFCP}$}
}
\caption{\textsc{MFCPM}\label{IR}}}
\end{algorithm}

 \linesnumbered
\begin{algorithm}[htbp]
 \footnotesize{
   \KwData{$\mathcal{FCP}_{k}$, \textit{minsupp}, and $\textit{min-all-confidence}$.} \KwResult{$\mathcal{FCP}_{k+1}$}
\Begin{
	 $\mathcal{FCP}_{k+1}$ := $\emptyset$\;
	 $\mathcal{C}_{k+1}$ := \textsc{Apriori-Gen}\textsc{(}$\mathcal{FCP}$$_{k}$\textsc{)}\label{generation_candidats}\;
	 \ForEach{\textsc{(}$X_{k+1}$ in $\mathcal{C}_{k+1}$\textsc{)}}
	 {
		   \If{\textsc{(}$\exists \ x, y \in X_{k+1}$ $|$ $\displaystyle\frac{\textit{Supp}\textsc{(}x\textsc{)}}
									{\textit{Supp}\textsc{(}y\textsc{)}} < \textit{min-all-confidence} $\textsc{)}}
		   {stop\label{test_crosssupp}; /* stop treatment of current candidate */\\}
\If{\textsc{(}$\exists \ X_{k} \subset X_{k+1}$ $|$ \textsc{(}$X_{k} \notin \mathcal{FCP}_{k}$\textsc{)}\textsc{)}\label{idealordre}}
				{
				stop; /* stop treatment of current candidate */\\

}
		   {			
			\If{\textsc{(}$X_{k+1}$.\textit{supp} $\geq$ \textit{minsupp}\textsc{)}\label{test_frequence}}
			{
			 $X_{k+1}.\textit{almax}$ := \textit{max}$\{\textit{Supp}\textsc{(}i\textsc{)} \ | \ i\in X_{k+1}\}$\;
			  $\textit{all-confidence}$\textsc{(}$X_{k+1}$\textsc{)} :=  $\displaystyle\frac{X_{k+1}.\textit{supp}}{X_{k+1}.\textit{almax}}$\label{calcul_alconfidence}\;
			  \If{\textsc{(}\textit{all-confidence}\textsc{(}$X_{k+1}$\textsc{)} $\geq$ $\textit{min-all-confidence}$\textsc{)}\label{test_correlation}}
			  { $\mathcal{FCP}_{k+1}$ := $\mathcal{FCP}_{k+1} \cup
X_{k+1}$\;
				}
			 }
			}
		  }
	
	\Return{$\mathcal{FCP}_{k+1}$}\label{resultat2}
 }
\caption{\textsc{Generate$\_$Next$\_\mathcal{FCP}$}\label{NextMFC}}}
\end{algorithm}

At the end of this data mining step, we obtain the patterns that will be used in the next step dedicated to replication.

\section{Replication strategy based on maximal frequent correlated pattern mining}\label{sec:page style}

In this section, we present our dynamic strategy, called RSCP, which is a periodic and decentralized replication strategy dedicated to the P2P topology. Noteworthily, this choice is argued by the fact that the P2P network is flexible in terms of communication between grid sites \cite{P2PvsGrid2003}. The proposed strategy considers the storage space capacity within each site as limited, contrary to a large part of periodic strategies. RSCP aims to improve grid performance through co-locating correlated files. It performs as follows:
\begin{itemize}
  \item[$\bullet$] \textbf{Step 1: \textit{Extracting and converting the file access history}}\\At each period of time quantified in number of executed jobs, each site keeps track of its access history carried out by the jobs executed on it for all local files and remote ones.
        A file access history describes then all files accessed by each job executed in the site. In addition, for each couple \textsc{(}job, file\textsc{)} an access number is assigned which means the number of times a job has accessed a given file during a period. The file access history is then converted into a binary context, \textit{i.e.}, a logical file access history. A binary context results in a table containing logical values where accessed files are considered as items, while the required files of each job constitute each transaction.

  \item[$\bullet$] \textbf{Step 2: \textit{Mining maximal frequent correlated patterns}}\\In this step, we apply a new maximal frequent correlated pattern mining algorithm in order to discover the hidden correlations between files. Each mined pattern then conveys information about closely related files, frequently used by jobs.

  \item[$\bullet$] \textbf{Step 3: \textit{Performing the replication and replacement processes}}\\Instead of considering files separately, the granularity used during this step consists in groups of correlated files. Maximal frequent patterns identified in the previous step constitute then the input of the replication algorithm. For each group of correlated files to be replicated, if there is enough storage space to hold all files of the group in the considered site, then the replication of these correlated files will be performed. Otherwise, replacement should be done. For this purpose, the candidate files for deletion are selected according to their weight. If the weight of the group of files to replicate is larger than the weight of the candidate files for deletion, then these latter ones are replaced by the files to replicate. Otherwise, the replication will not be carried out.
\end{itemize}


\subsection{Extracting and converting the file access history}
Periodically, each site maintains a file access history. Formally, we define the file access history on site $S_{i}$ as a matrix $\mathcal{A}$ of $n$ $\times$ $m$ integer values, where $n$ is the number of total jobs executed on the site $S_{i}$ during a given period and $m$ is the total number of files accessed by these jobs, such that $A_{j,k}$ = $\#Request_{J_{k},F_{j}}$, which is the number of times the job $J_{k}$ accessed the file $F_{j}$.\\

Before proceeding to data mining, the file access history must be converted to a binary context containing logical values \textsc{(}0 or 1\textsc{)}. 
For this purpose, we take into account the file popularity \cite{CC2016,JNCA2016}. A file $F_{j}$ is said popular within the site $S_{i}$ if the jobs executed in $S_{i}$ request frequently this file. We introduce the average number of requests for each file, denoted  $AvgAccess_{F_{j}}$, to evaluate the popularity of a file $F_{j}$ in a site $S_{i}$. It is defined as follows:
\begin{equation}\label{AvgAccess}
AvgAccess_{F_{j}} = \frac{\sum\limits_{k=1}^{n_j}\#Request_{J_{k},F_{j}}}{n_j}
\end{equation}
where $n_j$ represents the number of jobs executed in the site $S_{i}$ that access the file $F_{j}$.\\


Algorithm \ref{ConvertirCE} describes the step dedicated to the conversion of the file access history to a binary context. 
The theoretical complexity of this algorithm is in $O\textsc{(}m \times n\textsc{)}$ where $m$ is the number of accessed files by the $n$ jobs executed in the site.

\linesnumbered
\begin{algorithm}[htbp]
 \footnotesize{
 \KwData{File access history: $\{$\textsc{(}$J_{k}$, $F_{j}$, $\#Request_{J_{k},F_{j}}$\textsc{)} $|$ $k$ = $1\ldots n$, $j$ = $1\ldots m$$\}$, $n$ is the number of jobs executed in the site and $m$ is the number of accessed files.}
	\KwResult{Binary context: BinC}
\Begin{
\ForEach{file $F_{j}$} {
Compute $AvgAccess_{F_{j}}$\;
\ForEach{job  $ J_{k}$} {
\eIf{$\#Request_{J_{k},F_{j}} \geq AvgAccess_{F_{j}} $}{
		\textit{BinC$_{j,k}$} := 1\;
}{\textit{BinC$_{j,k}$} := 0\;}
}}
}
 \caption{\textsc{Converting$\_$to$\_$Binary$\_$Context}\label{ConvertirCE}}}
\end{algorithm}

\subsection{Mining the maximal frequent correlated pattern}
In this step, the proposed algorithm for mining the adopted concise representation in our work, namely the set of maximal frequent correlated patterns \textsc{(}\textit{cf.} Section \ref{sec:MFCP}\textsc{)}, is invoked. It is applied on the binary context obtained from the previous step. The used representation contains the smallest number of correlated patterns. Each retained pattern includes a maximal set of files frequently occurring and having a high correlation between them. Then, such a representation avoids retaining the whole set of correlated patterns, which is often very large and contains redundancy.

\subsection{Performing the replication and replacement processes}

In this step, the set $\mathcal{MFCP}$ of maximal frequent correlated patterns identified in the data mining step contains the candidates for replication. Each element $MFCP_{i}$ $\in$ $\mathcal{MFCP}$ is composed by files that are accessed frequently and simultaneously by jobs. Algorithm \ref{Rep} performs replication as follows:
\begin{itemize}
\item As shown in line 3, maximal frequent correlated patterns are sorted w.r.t. the    descending order of their size \textsc{(}\textit{i.e.}, the number of items in each pattern\textsc{)}.

\item For each $MFCP_{i}$ $\in$ $\mathcal{MFCP}$, if the free storage space is sufficient to store all the files contained in the correlated pattern, then the replication of these files will be carried out in $S_{i}$ \textsc{(}\textit{cf.} lines 4-5\textsc{)}.

\item Otherwise, a set of candidate files for deletion will be selected according to their weight \textsc{(}\textit{cf.} lines 6-16\textsc{)}.

In this regard, we define the file weight for each file $F_j$ w.r.t. a site $S_i$ as follows:
\begin{equation}
\textit{FileWeight}_{F_j} =  \frac{Size_{F_j} \times \#Request_{S_{i},F_j}}{BW_{S_{i},S_{r}}}
\end{equation}
where: $Size_{F_j}$ denotes the size of $F_j$, $BW_{S_{i},S_{r}}$ is the bandwidth between the site $S_{i}$ and the site $S_{r}$ containing the best
replica of $F_j$, and $\#Request_{S_{i},F_j} = \sum\limits_{k=1}^{n} \#Request_{J_{k},F_j}$ represents the
request number of $F_j$ by the site $S_{i}$, \textit{i.e.}, the number of times jobs $J_{k}$ executed in $S_{i}$ request the file $F_j$.\\Note that the best replica is the one that will be accessed as faster as possible based on the bandwidth parameter. It is chosen by examining the available bandwidth between the requesting site for a file and all sites containing a replica of the required file. Such an access will allow decreasing the job execution time.

Hence, according to the formula of the weight, unpopular files and files located in
the sites having a high bandwidth with $S_{i}$ are considered first for deletion.

\item Then, two average weights are calculated: the first for the group of files to replicate and is denoted $AvgGroupRepWeight$, while the second is associated to the selected group for deletion and is denoted $AvgCandidateDelWeight$ \textsc{(}\textit{cf.} lines 17-22\textsc{)}. Both associated formulae are as follows:

     \parbox{14.cm}{
         \hspace{0.5cm}
   $AvgGroupRepWeight$ = $\frac{\displaystyle
1}{\displaystyle
|ToReplicate|}$ $\times$ $\sum\limits_{f \in ToReplicate}^{} FileWeight_{f}$}

    \parbox{14.cm}{
    \hspace{0cm}
$AvgCandidateDelWeight$ = $\frac{\displaystyle
1}{\displaystyle |CandidateDel|}$ $\times$   $\sum\limits_{f \in CandidateDel}^{} FileWeight_{f}$}\mbox{}\\

\item Finally, if the first average is larger than the second one, candidate files for deletion are replaced by the new correlated group of files \textsc{(}\textit{cf.} lines 23-25\textsc{)}.
\end{itemize}
The theoretical complexity of this algorithm is in $O\textsc{(}k \ log \ k \ + \ k \times \ \textsc{(}p \ log \ p \ + \ n\textsc{)}\textsc{)}$ where $k$ represents the size of $\mathcal{MFCP}$, $p$ is equal to the number of files locally stored in $S_i$ when the strategy is invoked, and $n$ is the size of the longest maximal frequent correlated pattern contained in $\mathcal{MFCP}$.

\linesnumbered
\begin{algorithm}[!t]
\footnotesize{
    	\KwData{The set $\mathcal{MFCP}$ associated to a site $S_i$.}
    \KwResult{Correlated files replicated in $S_i$ or existing files retained instead.}
\Begin{
Sort the maximal frequent correlated patterns $MFCP_{i}$ $\in$ $\mathcal{MFCP}$
in descending order of size\;
\ForEach {group of files $MFCP_{i}$ $\in$ $\mathcal{MFCP}$} {
	\eIf{ \textsc{(}$ \sum\limits_{F_{j} \in MFCP_{i}, F_{j} \notin S_{i}}^{} Size_{F_j} \leq S_{i}.FreeSpace$\textsc{)}}{
Get each file $F_{j} \in MFCP_{i}, F_{j} \notin S_{i}$, from the site $S_{r}$ containing the best replica of $F_{j}$ and replicate it in $S_{i}$\;}
{Calculate FileWeight for each file in $S_{i}$ and sort files by FileWeight in ascending order\;
$CandidateDel$ := $\emptyset$\;
 AccumulativeFreeSpace := $S_{i}.FreeSpace$\;
AccumulativeCandDelWeight := 0\;
\While{\textsc{(}AccumulativeFreeSpace $<$
$\sum\limits_{F_{j} \in MFCP_{i}, F_{j} \notin S_{i}}^{} Size_{F_j}$\textsc{)}}{
 Get the first file $F_{candidate}$ from the sorted list\;
 AccumulativeFreeSpace += $Size_{F_{candidate}}$\;
AccumulativeCandDelWeight += $FileWeight_{F_{candidate}}$\;
$CandidateDel$ := $CandidateDel$ $\bigcup$ \{$F_{candidate}$\}\;
Delete $F_{candidate}$ from the sorted list\;
}
$AvgCandidateDelWeight$ := $\frac{AccumulativeCandDelWeight}{|CandidateDel|}$\;
$ToReplicate$ := $\emptyset$\;
\ForEach{file $ F_{j} \in MFCP_{i}, F_{j} \notin S_{i} $} {
GroupRepWeight += $FileWeight_{F_{j}}$\;
$ToReplicate$ := $ToReplicate$ $\bigcup$ \{$F_{j}$\}\;
}
$AvgGroupRepWeight$ := $\frac{GroupRepWeight}{|ToReplicate|}$\;
\If{$AvgCandidateDelWeight$ $\leq$ $AvgGroupRepWeight$}{
Delete files of $CandidateDel$ from $S_{i}$\;
Replicate correlated files of $ToReplicate$ to $S_{i}$\;
}}}}
\caption{\textsc{Replication process}\label{Rep}}}
\end{algorithm}

\section{Performance evaluation}

\subsection{Experiment environment}

We concentrate here on some of the representative results obtained. Experiments were carried out using the OptorSim simulator \cite{CameronEtAlOptorSim} while considering four access patterns, namely sequential, random, random walk Gaussian, and random Zipf, for a fixed number of jobs equal to 1 000. Further details on the experiment environment and the obtained experimental results can be found in \cite{JSS2015,PDCAT2014}.

\subsection{Analysis of the obtained experimental results}

RSCP is compared with four other replication strategies: No Replication \cite{Ranganathan01identifyingdynamic}, DR2 \cite{DR2}, PRA \cite{Preftching08} and PDDRA \cite{PDDRA12}. 
We point out that performance comparison of our strategy with No Replication is required because any strategy that performs worse than No Replication is not worth considering
\cite{Ranganathan01identifyingdynamic}.\\

The obtained results w.r.t. the different used metrics are as follows:
\begin{itemize}
\item \textbf{Mean job time:} The mean job time is defined as the total time in \textit{ms} of all the jobs divided by the number of jobs completed. 
Figure \ref{mjt} illustrates the mean job execution time for the five strategies No Replication, DR2, PRA, PDDRA and RSCP for the different access patterns. The simulation results show that RSCP has the lowest value of the mean job execution time with most access patterns, namely sequential access, random access and random walk Gaussian access. This can be explained by the fact that if the correlated files, jointly requested by a set of jobs, are co-located, \textit{i.e.}, stored in the same site running them, these jobs will have their required files locally. Hence, the number of local accesses will increase and by the way the number of remote accesses will decrease. Thus, the total job execution time will be minimized since the data transfer will be reduced as much as possible. In the Zipf access pattern, PDDRA outperforms our strategy. This can be argued by the fact that, once the Zipf access pattern is applied, only few files are accessed frequently while several others are rarely accessed. In this case, looking for high correlated sets of files through a data mining algorithm is not beneficial since the number of frequent files is reduced.

\begin{figure}[h]
\centering
\includegraphics[scale=0.8]{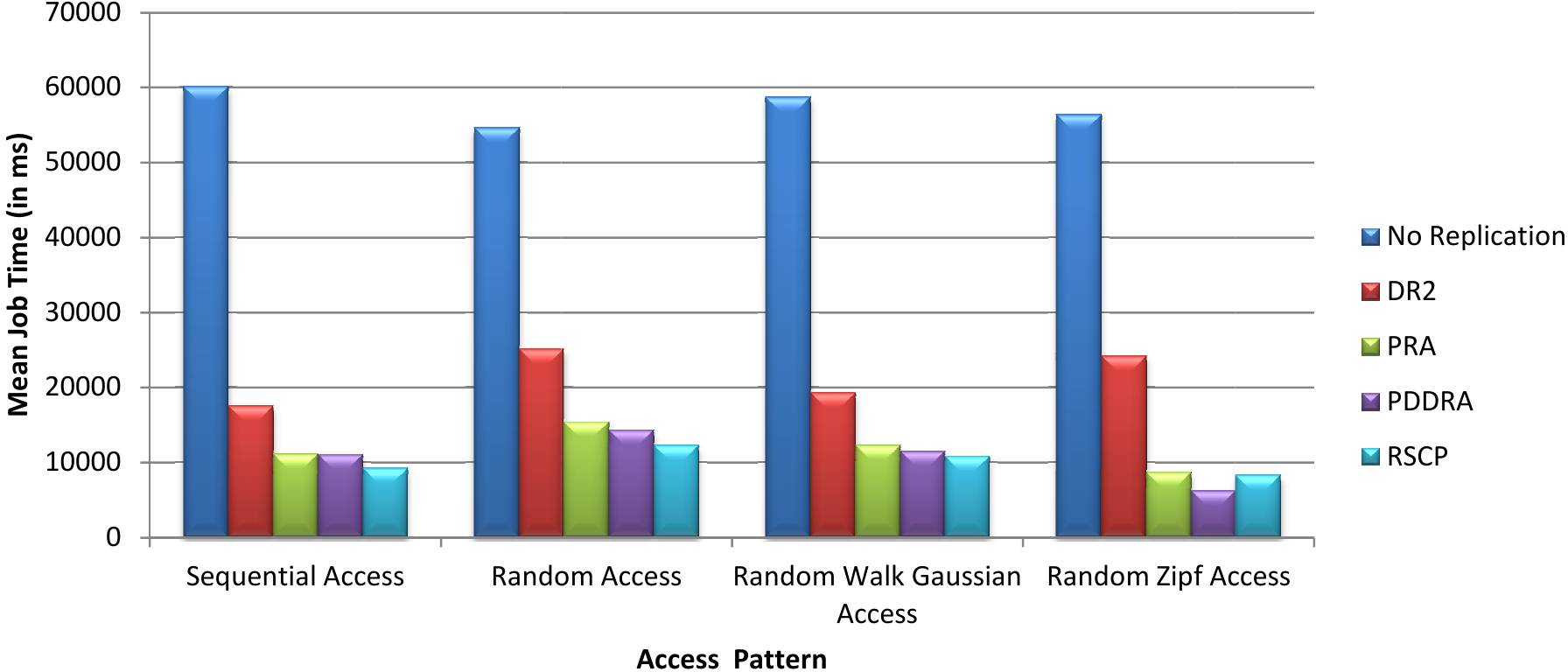}
\caption{Mean job time for different access patterns}\label{mjt}
\end{figure}

\item \textbf{Effective Network Usage \textsc{(}ENU\textsc{)}:} Figure \ref{enu} shows the results of the ENU evaluation for the five strategies. Again our strategy has the lowest value of ENU with most access patterns in comparison with the other strategies. The main reason is that jobs will locally have their required files at the time of need. Hence, remote accesses will decrease and the total number of local accesses will increase. Among its main aims, our strategy has to minimize bandwidth consumption and thus it allows decreasing the network traffic.

\begin{figure}[h]
\centering
\includegraphics[scale=0.8]{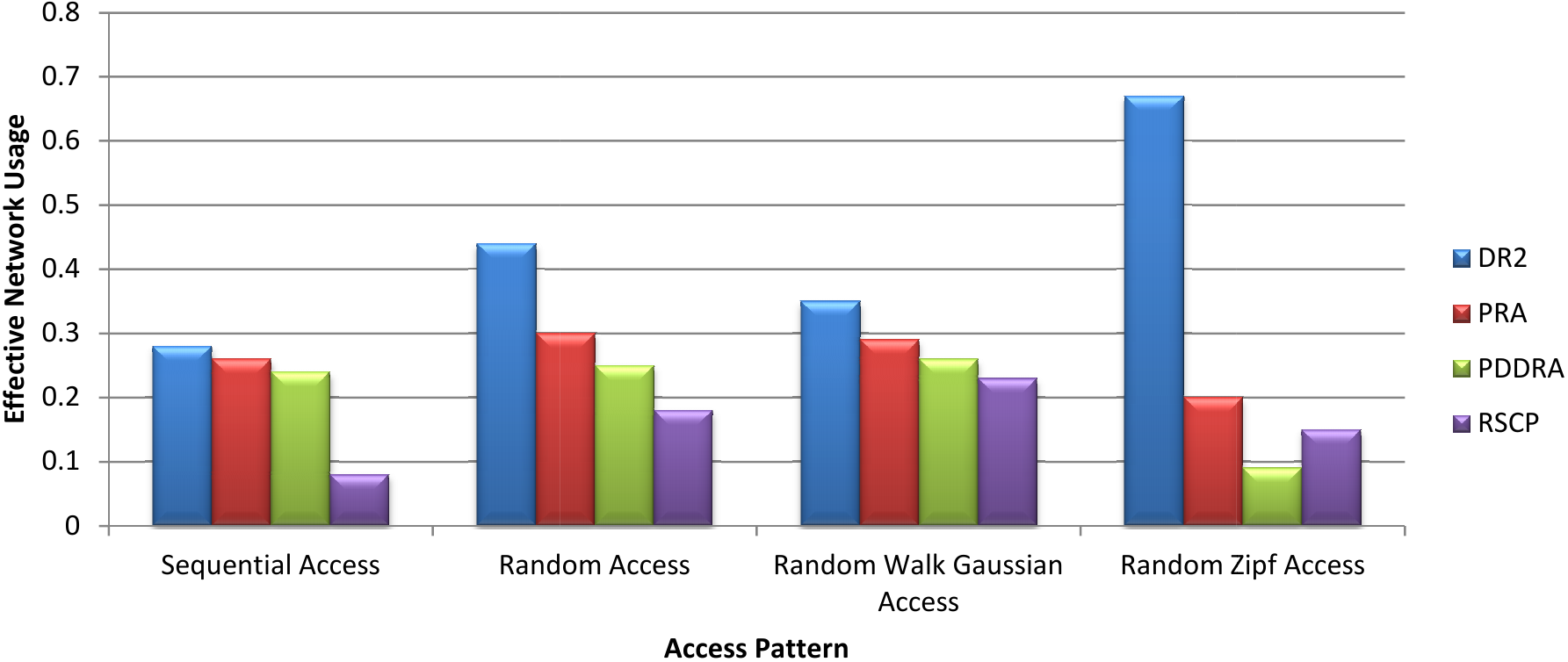}
\caption{Effective network usage for different access patterns}\label{enu}
\end{figure}

\item \textbf{Total number of replications:} As far as this number is low, as much is the efficiency of the strategy from the view point of this metric since with more number of replicas, the cost of maintaining them becomes an overhead for the system. 
Figure \ref{nr} and Table \ref{tnr} show the total number of file replications performed by the tested strategies. Here, with all access patterns, our strategy gives the lowest value of number of replications in comparison with the other strategies while ensuring a good level of files availability in the data grid. This can be explained by the fact that patterns selected as a candidate to be replicated must simultaneously fulfill both frequency and correlation constraints. Hence, the number of replicas created is more reduced and more accurate than those obtained with the other replication strategies.
    We can also remark that the number of replications carried out by the PRA and PDDRA strategies is very high. This is because both these strategies perform replication and predict future requests at each file request. As aforementioned, a higher number of replications means a great number of file transmissions. Thus, these strategies consume a considerable amount of network bandwidth. 

\begin{table}[htbp]
\centering
\footnotesize \begin{tabular}{|p{4.6cm}||p{2.3cm}|p{0.8cm}|p{0.8cm}|p{1.2cm}|p{1.5cm}|}
	\hline
	
	\makebox[\linewidth][c]{\textbf{Access pattern}}& \makebox[\linewidth][c]{\textbf{No Replication}} & \makebox[\linewidth][c]{\textbf{DR2}} &\makebox[\linewidth][c]{\textbf{PRA}}&\makebox[\linewidth][c]{\textbf{PDDRA}}&  \makebox[\linewidth][c]{\textbf{RSCP}}\\

	\hline  \hline
	
\makebox[\linewidth][c]{Sequential Access}&\makebox[\linewidth][c]{0} &\makebox[\linewidth][c]{197}&\makebox[\linewidth][c]{494}&
\makebox[\linewidth][c]{522}&\makebox[\linewidth][c]{159}  \\ \hline
\makebox[\linewidth][c]{Random Access} &\makebox[\linewidth][c]{0}&
\makebox[\linewidth][c]{118} &\makebox[\linewidth][c]{360}&\makebox[\linewidth][c]{389}&
\makebox[\linewidth][c]{102} \\ \hline
\makebox[\linewidth][c]{Random Walk Gaussian Access} &
\makebox[\linewidth][c]{0} &
\makebox[\linewidth][c]{164}&
\makebox[\linewidth][c]{348}&
\makebox[\linewidth][c]{356}&
\makebox[\linewidth][c]{108} \\ \hline
\makebox[\linewidth][c]{Random Zipf Access} &\makebox[\linewidth][c]{0} &\makebox[\linewidth][c]
{29}&
\makebox[\linewidth][c]{116}&\makebox[\linewidth][c]{343}&\makebox[\linewidth][c]{79} \\ \hline
	   \end{tabular}
\caption{Total number of replications for different access patterns} \label{tnr}
\end{table}

\begin{figure}[h]
\centering
\includegraphics[scale=0.8]{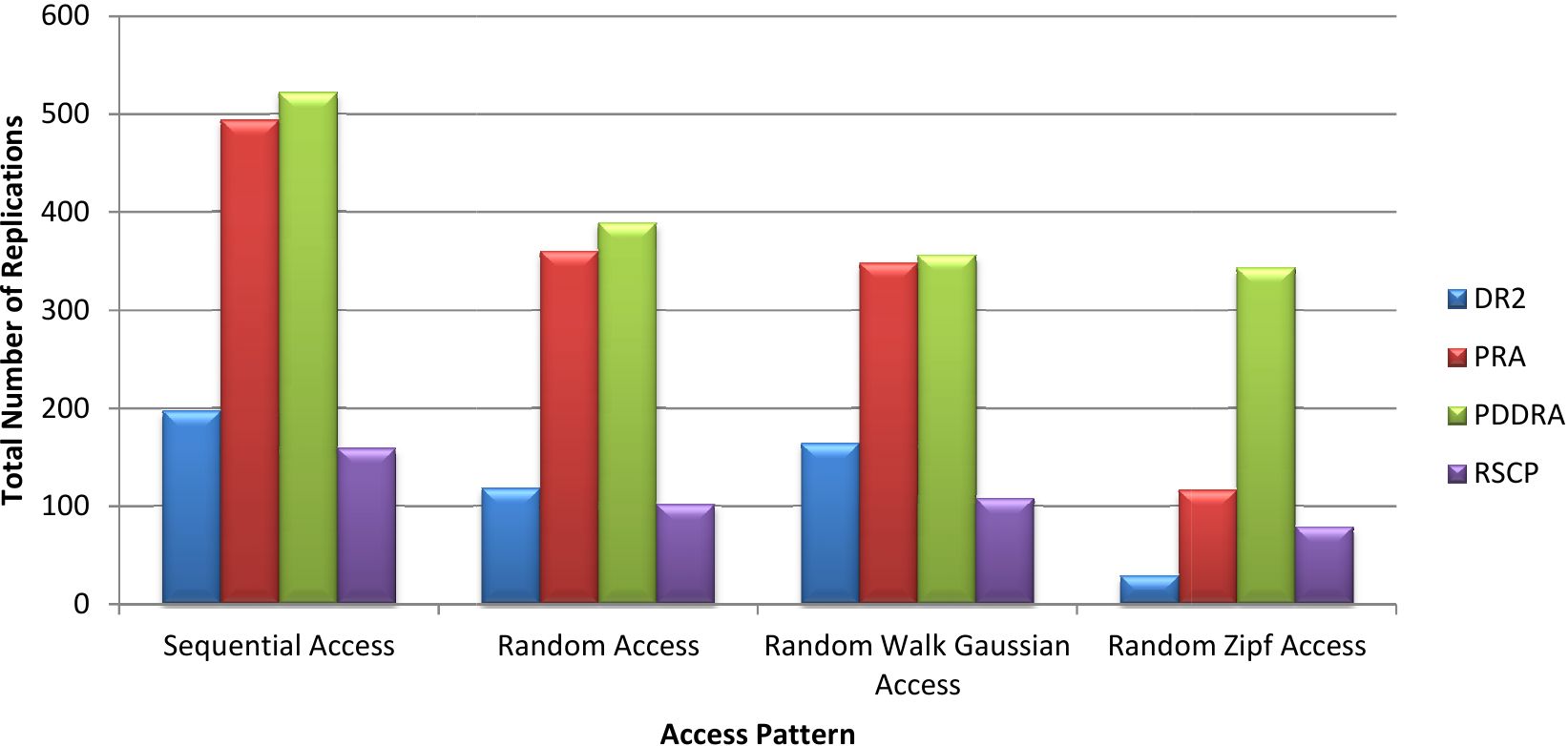}
\caption{Total number of replications for different access patterns}
\label{nr}
\end{figure}

\item \textbf{Hit ratio:} The hit ratio computes the ratio of the total number of local file accesses to all accesses containing the local file accesses, the total number of file replications and the total number of remote file accesses as follows: $$\frac{N_{local\ file\ accesses}}{N_{remote\ file\ accesses} + N_{file\ replications} + N_{local\ file\ accesses}}$$
    Hit ratio ranges from 0 to 1 as the ENU parameter. However, a high value of
this performance metric is preferable. In order to evaluate this metric, let us consider Table \ref{LRaccess} which shows the numbers of local and remote file accesses of each strategy for different access patterns, as well as Table \ref{tnr} which shows the total number of replications. In Table \ref{LRaccess}, NLFA is the number of local file accesses and NRFA is the number of the remote file accesses. As we can see, No Replication does not perform any local file access. For this reason, the NLFA value of this strategy is equal to 0 with all access patterns. Figure \ref{hr} shows the hit ratio evaluation for the five strategies. We can remark that our strategy has the greatest value of hit ratio for almost all access patterns.

\begin{table}[h]
\begin{center}
\twlrm
\begin{tabular}{|c||c|c||c|c||c|c||c|c|}
\hline
\multirow{3}{*}{\bf Strategy}  & \multicolumn{2}{c||}{\bf Sequential Access} & \multicolumn{2}{c||}{\bf Random Access} & \multicolumn{2}{c||}{\bf Random Walk}& \multicolumn{2}{c|}{\bf Random Zipf}\\

& \multicolumn{2}{c||}{}&\multicolumn{2}{c||}{}& \multicolumn{2}{c||}{\bf Gaussian Access}& \multicolumn{2}{c|}{\bf Access}\\

 \cline{2-9}
& NLFA & NRFA  &NLFA & NRFA  &NLFA & NRFA &NLFA & NRFA\\
\hline \hline
No Replication  & 0 & 12546 & 0 & 12386 & 0 & 12360& 0 & 12488\\ \hline
DR2             &9177& 3382 & 7022& 5521 & 8130& 4282 &4032& 8171\\ \hline
PRA             &9818& 2856 & 8983& 3354 & 9301& 3363 &9839& 2438\\ \hline
PDDRA           & 9746& 2490&9592& 2708&10118& 3188&11189& 792\\ \hline
RSCP            & 8646& 594& 7749& 1634&7353& 2081&8047& 1415\\ \hline
\end{tabular}
\caption{Number of local and remote file accesses for different access patterns}\label{LRaccess}
\end{center}
\end{table}

\begin{figure}[!t]
\centering
\includegraphics[scale=0.8]{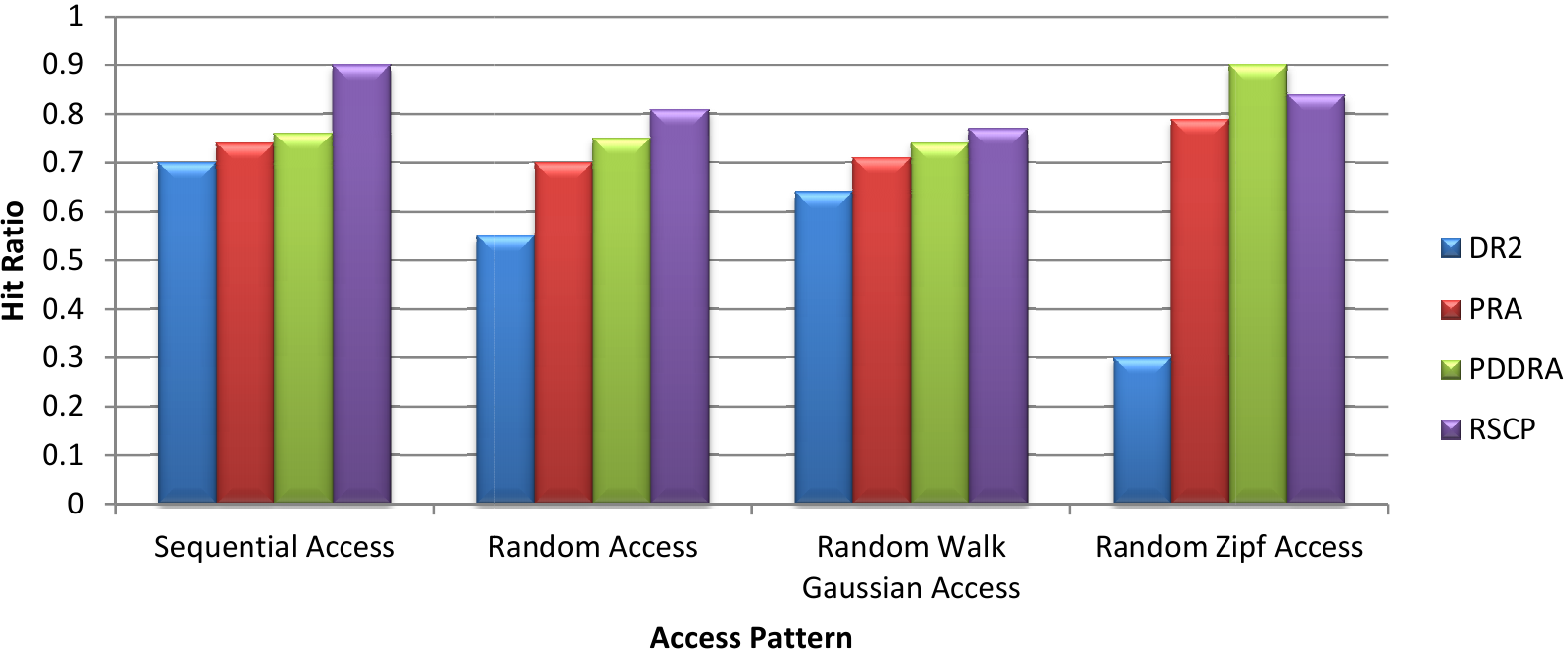}
\caption{Hit ratio for different access patterns}\label{hr}
\end{figure}

\item \textbf{Percentage of storage filled:} The percentage of storage filled is equal to the average percentage of capacity in Mb of the storage elements in all grid sites used by files. A strategy is considered efficient w.r.t. this metric if it performs a small number of replications which leads to a low percentage of storage filled. Figure \ref{ps} shows the percentage of storage filled evaluation. The No replication strategy gives the minimum percentage of storage used since it does not perform any replication at all. In addition, we can observe that with all access patterns, our strategy has the lowest value in comparison with the other replication strategies since it allows to create new replicas only when this is beneficial.

\begin{figure}[!t]
\centering
\includegraphics[scale=0.8]{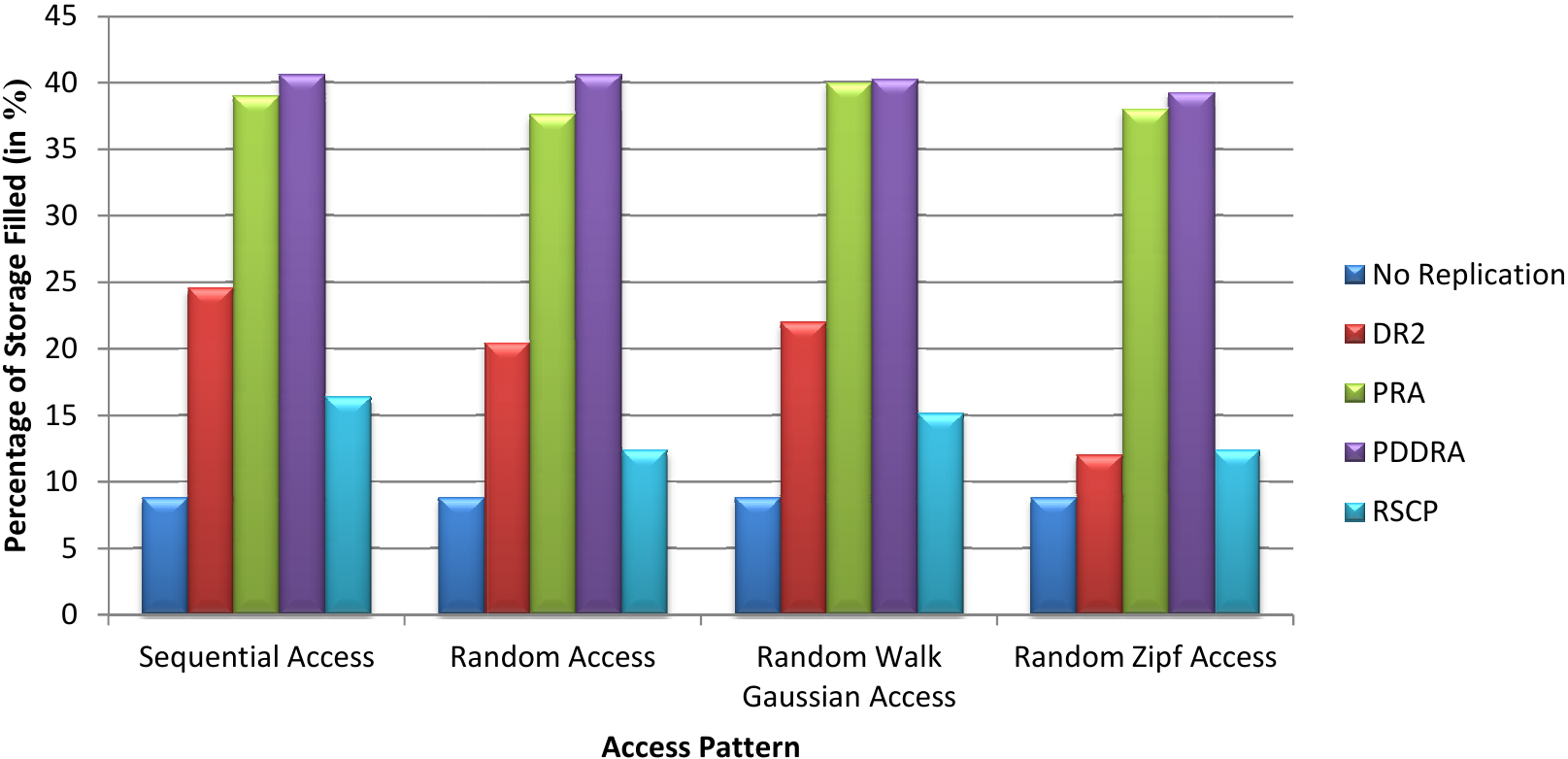}
\caption{Percentage of storage filled for different access patterns}\label{ps}
\end{figure}
\end{itemize}

%

\section{Conclusion}
In this chapter, a new replication strategy called RSCP is proposed. RSCP, which is executed periodically at each site, takes into account correlations between files and considers groups of correlated files as granularity for replication. Its main goal is to co-locate files that are frequently and simultaneously accessed by jobs in the same site. When storage space is insufficient to replicate all members of a group of correlated files, the strategy selects a group of locally available files for deletion. Selection is based on a dedicated metric noted weight which considers both file and network parameters. Using several metrics and access patterns, carried out experiments prove the efficiency of RSCP compared to existing replication strategies.


\chapter*{Conclusion and Future Work}
\addcontentsline{toc}{chapter}{Conclusion and Future Work}
\markboth{Conclusion and Future Work}{Conclusion and Future Work}
\setcounter{footnote}{0}

In this report, we presented a synthesis of our main contributions in the field of replication in data grids. These contributions were divided into two axes according to the target goals:
\begin{itemize}
  \item The first axis deals with the evaluation of the distribution quality of replicas over grid sites. Indeed, through the replication technique, a file has several replicas distributed on different sites according to some criteria on which the associated placement step is designed. As a consequence, different replication strategies generate distinct replicas distributions. Each induced distribution will then affect the performance of future uses of the grid. In this situation, most of the metrics used in the literature to evaluate replication strategies do not take into consideration the initial replicas distribution on which a strategy is invoked although the distribution quality strongly affects the obtained metrics results.

       In this context, how to quantify the quality of a distribution is a key task. Indeed, it offers key information on the effect of a strategy on replicas distribution. On the other side, offering the possibility to quantify the distribution quality helps to predict the quality of service a grid can offer to according to the resulting data distribution. This hence allows to progressively adapt the behavior of the grid and to transit then if necessary from a strategy to another. 

      We then proposed several evaluation metrics dedicated to the assessment of the distribution quality. Each metric has its proper characteristics and was designed on the basis of some criteria. The interaction between a replication strategy and the distribution of replicas on sites was hence analyzed and the influence of each one -- the strategy or the distribution -- on the other was studied. Furthermore, assessing the distribution quality of data made it possible setting up an objective evaluation of replication strategies.

  \item The second axis we explored is based on the synergy highlighted in the literature between data mining and data grid. Indeed, data grids offer an adequate infrastructure for running data mining algorithms on highly sized volume of data. While data mining techniques constitute interesting solutions for the analysis of the data gathered during the running of data grids. For example, the file access history offers information on files often accessed together by jobs or sites.

  In this respect, we highlighted the importance of data mining towards the enhancement of replication strategies performances. A survey of the main data mining based replication strategies, which is lacking in the literature to the best of our knowledge, was then proposed as well as a complete guideline helping researchers setting new replication strategies based on data mining techniques \cite{EAAI2016}. In addition,
  our work on concise representations of patterns \cite{IDA2012,IJAIT2014} and more particularly correlated patterns \cite{DS2010,IDA2015} was at the roots of the design of a new replication strategy, based on the application of correlated patterns in the context of data grids. Indeed, although some already existing strategies of the literature rely on pattern mining, they suffer from the quality of the mined patterns after the data mining step as well as their high number. The integration of a correlation measure offers more informative patterns while reducing the number of mined patterns. In addition, aiming at going beyond the offered size of the whole set of correlated patterns, we proved the utility of only mining maximal frequent ones. This allows a shrewd reduction of the patterns to be used in the replication step while retaining the maximal sets of files simultaneously requested, in a frequent manner.
\end{itemize}

Obtained theoretical results are consolidated by various experiments carried out using the OptorSim simulator, and in which the proposed metrics and strategies were implemented, tested and compared to those of the literature.\\


The obtained results open some short-term as well as some long-term perspectives which are related to the following directions.

\subsection*{Evaluation of the distribution quality: towards more sophisticated metrics and extended applications}
\begin{enumerate}
   \item The design of new hybrid evaluation methods which combine the key properties of those already proposed in this work. It will be interesting in this respect to set up a synergy with research looking for determining the optimal number of replicas a data grid should contain \cite{mansouriQoS}. This will ensure a more optimized use of the data grid. In addition, other grid parameters like replica size, storage capacity and cost, workload, etc., should be taken into consideration in order to design more sophisticated metrics dedicated to the assessment of the distribution quality. Coupled with new correction processes, a more precise correction of the results of evaluation metrics can be offered while ensuring that the error rate associated to the assessment process of the corrected values does not exceed a pre-defined threshold.

   \item The design of new replication strategies having for main task to improve the quality of replicas distribution is also interesting. For such a purpose, the criterion of the decision to make a replication will be the ability or not to improve the quality of the distribution according to a given dedicated metric. In case of insufficient available space, misplaced files w.r.t. the distribution quality metric will be deleted.

       In the same context, when placing a new replica of a file, several methods can be applied in order to ensure a high distribution quality level. First results we obtained are encouraging. Indeed, in \cite{CoopIS2016}, we rely on a barycenter-based approach while considering a given weight for each site not containing the file to be replicated. The best location -- one of the considered sites -- is then selected according to the barycenter properties. This satisfies the needs of a set of sites instead of mainly only one, and contributes therefore in the improvement of the distribution quality of replicas.

       In this respect, how to correctly quantify the popularity of a given replica, w.r.t. the requesting sites, plays a key role in the design of efficient strategies as highlighted in our work done in \cite{CC2016,JNCA2016}. In addition, the availability of sites should be precisely determined since it plays an important role in the choice of where to place replicas as well as where to execute a job \cite{SiteAvailability2015}.

\item From an experimental point of view, we plan to extend the carried out study to a large set of replication strategies as well as to different evaluation metrics. It may then be an interesting research opportunity to conduct an experimental study to compare several existing data replication strategies, to show the ones affected less by the initial distribution.

   \item Extending the evaluation of the distribution quality to a set of objects \textsc{(}replicas, services, etc.\textsc{)} over system locations for different distributed systems, like clouds, constitutes a promising issue.\\
 \end{enumerate}

\subsection*{Use of data mining techniques in replication strategies: targeting high-quality patterns and going further in their exploitation}
\begin{enumerate}
\item The current version of the RSCP strategy offers several possibilities of extension as follows:
    \begin{enumerate}
       \item The data on which the data mining technique will be invoked have a paramount importance w.r.t. the quality of the mined patterns. On the one hand, several other discretization processes can be used to convert the file access history into a binary context. On the other hand, we may also design a dedicated algorithm for quantitative correlated pattern mining 
           in order to apply it directly on the file access history containing numerical values. A comparison can then be made between approaches starting from a Boolean context and those from a numerical one w.r.t. the mining complexity versus the quality of the mined patterns.

      \item At each period, the data mining step can give several maximal frequent correlated patterns. Since a given file may belong to more than one patterns, some redundancy in the treatment occurs \textsc{(}like the computation of a file weight\textsc{)}. It will then be interesting to reduce as much as possible this overhead through reducing the number of mined patterns by only considering a set of top-k selected patterns according to some criteria. Moreover, in each pattern, files should be sorted according to an importance value to be considered. This will allow to effectively replicate a subset of the files if there is not sufficient space for storing the whole pattern files. In this situation, a centralized analysis of the mined patterns in the different sites can further help in offering an optimized number of replicas per file for the whole grid. Indeed, when replicating a file at a given site, already existing replicas of this file in its neighboring sites will be taken into consideration.

      \item The concise representation of correlated patterns used in the current version offers the advantage of being of reduced size. However, it is an approximate one in the sense that it does not allow to determine accurately the support and the all-confidence values of a given frequent correlated pattern which is not part of the representation \textsc{(}\textit{i.e.}, which is not maximal\textsc{)}. In case where such information is necessary for each pattern, other exact concise representations can be used, like frequent closed correlated patterns \cite{DS2010}, while being of larger size. Furthermore, other correlation measures can be used like for example \textit{bond} having the same structural properties as all-confidence while being more restrictive. In the general case, a trade-off is then to be found between the accuracy and the size of the representation according to the requirements to be satisfied by the strategy.
                \end{enumerate}

\item In data grid, new transactions \textsc{(}jobs\textsc{)} as well as items \textsc{(}files\textsc{)} are continuously added to the database as time advances. In other words, file access histories are incremental in nature. Hence, it is then suitable to apply incremental and time series data mining techniques \cite{FuTimeSeries}. These latter ones allow taking into consideration continuous updates that may arises on the data to be mined. This offers the opportunity to simultaneously consider the file access history of the current period and those of the previous periods. We deem that it results in a better file correlations evaluation and it is more suitable to the dynamic nature of the grid.

\item Mining rare correlated patterns \cite{IDA2015} can offer interesting knowledge in different situations \cite{DBLP:conf/waim/YuCD07}. Once applied in the context of replication, such patterns may for example indicate the set of files to be deleted in case of no available space. They indeed offer the possibility to locate information that a classical approach does not make possible. On their side, disjunctive patterns also offer knowledge about alternative situations \cite{DKE09,IJAIT2014}. They can hence be applied for example in order to find a complementary set of sites in which a file can be replicated.\\
\end{enumerate}

\subsection*{Considering the general context of replication strategies}
\begin{enumerate}
\item In this work we mainly considered read-only data. However, in practice data are updatable. Hence, maintaining the consistency of the data replicated across grid sites is a complicated issue that must be addressed \cite{consistency2017CCPE}.
    In particular, the evaluation process of the distribution quality of replicas should then be able to cover data grids that allow both read-only and write accesses. 

\item The adaptation of strategies initially proposed for data grids to be used on clouds and vice-versa seems to be an interesting issue. Indeed, grids and clouds offer several common issues \cite{FosterCloudvsGrid,VillegasCCBook2010}. For example, while taking into consideration the cloud properties, a data mining based technique can help in tackling the important task of discovering service dependencies w.r.t. a set of applications \cite{Popa:2009:MEA:1658939.1658966}.

    In this context, the use of algorithms for graph mining in order to mine clusters of sites or services that have common features seems to be a promising issue. Indeed, such techniques have been proved to be very efficient for solving problems that can be modeled using graphs like those related to social networks. In this respect, in several situations, system sites \textsc{(}or data center in the case of cloud\textsc{)} can be considered as nodes of a graph while links between sites can be regarded as edges.

\item It would be interesting to combine, through data mining techniques, data replication and job scheduling to improve the performance of grids. Indeed, as highlighted in several works \cite{simulatenousSchedulingReplicationRef1,simulatenousSchedulingReplicationRef2,simulatenousSchedulingReplicationRef3}, job scheduling and data replication are two problems which must be jointly studied. A request can be satisfied in reasonable deadline only if the data necessary for its execution are available on the one hand, and if there is an available site to process it on the other hand.

    In grids, we can consider a three-dimensional extraction context, called triadic context \cite{TriadicContext}, instead of a bi-dimensional context as used in the RSCP strategy and the other strategies of the literature based on data mining. This offers indeed a context-specific data in order to infer semantic relationships between files, sites and jobs, namely which file is requested from which site, by which job. Triadic concepts will be mined starting from such a context. These patterns will be used in the replication process of sets of correlated files while taking into consideration not only the relation between jobs and files they use but also between jobs and sites where they executed. This necessarily improves the scheduling process by selecting for each job the set of sites already containing a large part of the required files to not say the whole. Furthermore, using clustering techniques \cite{CloudPlacement2017}, such a triadic context offers the possibility to mine communities of sites, \textit{i.e.}, those sites interested in the same set of files/jobs. This can offer the possibility to partition the grid into several logical regions. In this respect, designing an extension of correlation measures in order to take into account a three-dimensional context constitutes a key step in this proposal.

    Moreover, triadic concepts are also applicable in the context of clouds in order to simultaneously satisfy replicating data near services or applications, \textit{i.e.}, data-oriented replication as well as replicating services closer to their client, \textit{i.e.}, service-oriented replication. Such a generic replication approach, simultaneously replicating data and services, allows minimizing the overall time required for executing services as well as the overall delivery time of all data files to their dependent services/application.

\item Further work should be dedicated to the consideration of more factors in exploring file correlations and not only the history of file accesses. In this case, using decision trees for example, prediction models of the value of a file attribute \textsc{(}like the popularity of a file\textsc{)} according to the other attributes \textsc{(}like its type, size, user, etc.\textsc{)} can be developed. The predicted value can be used to decide or not to replicate a file. In the general case, it would also be desirable to exploit data mining techniques in the resource prediction of the system needs.

\item From an experimental point of view, almost all replication strategies use simulation to evaluate and test the replication algorithms. Hence, it would be very interesting as a next step to test strategies in a real environment and even to confront results obtained using different simulators \cite{surveySimulationTools2013} \textsc{(}for example OptorSim \cite{CameronEtAlOptorSim} \textit{vs.} GridSim \cite{GridSim2002}\textsc{)}. In addition, it has been observed that most of strategies compare their results with some basic ones. Hence a lot of experiments are still required for thoroughly assessing their performances. Indeed, a quantitative study of these strategies through re-implementing them must be performed. This will allow setting an open-access platform for replication strategies implementations that can be extended through researchers aiming at making their implementations easily available for interested users. This will also offer a fair comparison between different approaches. Indeed, the ambiguity in the description of some strategies and the lack of necessary and somehow important details \textsc{(}for example, the absence of pseudo-code algorithms or the values of the used thresholds\textsc{)} make their implementation by others very difficult, not to say impossible. Furthermore, to go beyond simulations, studying the impact of replication strategies on real-life applications is an important issue that should be carried out in-depth. A first step towards this goal is to evaluate if the currently used simulators represent efficient and realistic benchmark tools or not. This consists in studying whether the performance evaluation using these tools is realistic.

In this same context, most of the strategies of the literature do not study the influence of different considered parameters on performance evaluation results. Indeed, parameters such as the period or in other words the history length \textsc{(}a small history window or a long one\textsc{)} as well as the used parameter thresholds are very important and have a great impact on the obtained results. Furthermore, for the case of data mining based replication strategies, since the execution of a replication strategy based on data mining technique closely depends on the performance of the used data mining algorithm, the overhead of the mining algorithm on the strategy performances should be assessed.
\end{enumerate}

\newpage

\addcontentsline{toc}{chapter}{Bibliography}
\bibliographystyle{splncs03}
\bibliography{UH_report}
\end{document}